\documentclass[prx,twocolumn,superscriptaddress,nofootinbib]{revtex4-2}
\usepackage{graphicx}
\usepackage{amssymb}
\usepackage{bm}
\usepackage{natbib}
\usepackage{amsmath}
\usepackage{mathrsfs} 
\usepackage{amstext}
\usepackage[english]{babel}
\usepackage[table, svgnames, dvipsnames]{xcolor}
\usepackage[colorlinks=true,citecolor=Cerulean,linkcolor=RubineRed,urlcolor=Cerulean]{hyperref}
\newcommand{\be}{\begin{eqnarray}}
\newcommand{\ee}{\end{eqnarray}}
\newcommand{\no}{\nonumber}

\begin{document}
\title{Shortcuts to Adiabaticity in Krylov Space}

\author{Kazutaka Takahashi\href{https://orcid.org/0000-0001-7321-2571} {\includegraphics[scale=0.05]{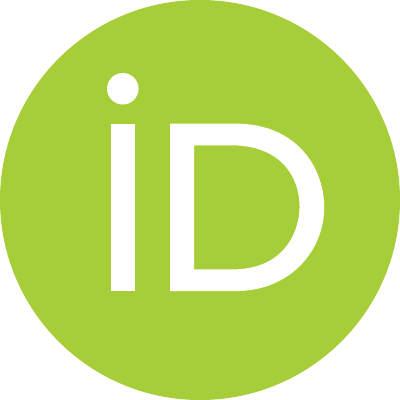}}}
\email{ktaka@phen.mie-u.ac.jp}
\affiliation{Department of Physics Engineering, Faculty of
Engineering, Mie University, Mie 514–8507, Japan}
\address{Department of Physics and Materials Science, University of Luxembourg,
L-1511 Luxembourg, Luxembourg}
\author{Adolfo del Campo\href{https://orcid.org/0000-0003-2219-2851}{\includegraphics[scale=0.05]{orcidid.pdf}}}
\email{adolfo.delcampo@uni.lu}
\address{Department of Physics and Materials Science, University of Luxembourg,
L-1511 Luxembourg, Luxembourg}
\address{Donostia International Physics Center, E-20018 San Sebasti\'an, Spain}

\begin{abstract}
Shortcuts to adiabaticity provide fast protocols 
for quantum state preparation in which the use of auxiliary counterdiabatic 
controls circumvents the requirement of slow driving in adiabatic strategies. 
While their development is well established in simple systems, 
their engineering and implementation are challenging in many-body quantum systems 
with many degrees of freedom. 
We show that the equation for the counterdiabatic term 
-- equivalently, the adiabatic gauge potential -- is solved by 
introducing a Krylov basis. 
The Krylov basis spans the minimal operator subspace 
in which the dynamics unfolds and provides an efficient way 
to construct the counterdiabatic term. 
We apply our strategy to paradigmatic single- and many-particle models. 
The properties of the counterdiabatic term are 
reflected in the Lanczos coefficients obtained in the course of 
the construction of the Krylov basis by an algorithmic method.
We examine how the expansion in the Krylov basis incorporates 
many-body interactions in the counterdiabatic term.
\end{abstract}

\maketitle

\section{Introduction}
In noisy quantum devices, dominant in the noisy intermediate-scale quantum (NISQ) era \cite{Preskill18},  the prospects of implementing exact adiabatic control protocols are dim. 
Noise generally lowers the fidelity of preparing a target quantum state, making the dynamics not unitary, and leading to a final mixed state. The presence of noise further limits the admissible operation time in adiabatic protocols, e.g., in adiabatic quantum computing and quantum annealing. In these devices, noise can act as a heating source leading to excitation formation \cite{Dutta16,Weinberg20}, precluding the goal of finding the low-energy configuration of a given problem Hamiltonian.

The ubiquitous presence of noise in current NISQ devices forces us to rethink the use of adiabatic strategies.
A natural approach is operating in timescales where environmental noise is negligible. A  demonstration of this approach has recently been reported in quantum annealing devices, where noise-induced errors generated for moderate operation times \cite{Bando20} can be eliminated by shortening the duration of the process \cite{King22}. However, this strategy generally limits the efficiency of the computation as a result of the adiabatic theorem, whether one considers the system closed \cite{Kato50} or open \cite{Venuti16}.
An alternative approach relies on optimally tailoring the time dependence of the parameters that are varied in time in the system of interest (e.g., the harmonic frequency in a trapped system or a magnetic field in a spin system) \cite{Garrido62}. This is the principle behind the so-called boundary cancellation method that reduces excitations by devising smooth protocols in view of the adiabatic theorem in either isolated or open systems \cite{Lidar09,AlbashLidar15,Venuti16}. Such an approach requires no additional control fields, easing the implementation of the driving protocols in the laboratory \cite{MunozBauza22}, but provides limited advantages in the speedup, and technical assumptions in the adiabatic theorem may restrict its applicability. 

Shortcuts to adiabaticity (STAs) provide an alternative approach \cite{Chen10,Torrontegui13,delcampo19,GueryOdelin19}. They enforce the nonadiabatic following of a prescribed adiabatic trajectory of interest, tailoring nonadiabatic excitations utilizing auxiliary control fields. In other words, STAs remove the requirement for slow driving in adiabatic protocols,  leading to the preparation of the same target state in a shorter time. 
By now, several experiments have demonstrated the use of STAs in ultracold atoms \cite{Schaff10,Schaff11,Schaff11njp,Bason12,Rohringer2015,Deng18,Deng18Sci,Diao18}, nitrogen-vacancy centers \cite{Zhang13,Kolbl19}, trapped ions \cite{An16}, and superconducting qubits \cite{Paraoanu19}, among others.

While various techniques have been developed to engineer STAs, counterdiabatic (CD) driving stands out among them by providing a universal approach for any system in isolation. 
The early formulation due to Demirplak and Rice \cite{Demirplak03,Demirplak05,Demirplak08}, independently developed by Berry \cite{Berry09}, assumes the dynamics to be unitary and the system Hamiltonian to be diagonalizable at all times.
However, progress over the past decade has shown that STAs can be applied to open quantum systems \cite{Vacanti14,Dann19,Dupays20,Alipour20,Funo21}, as demonstrated in a pioneering experiment \cite{Yin22}.

From the outset, the need for the system Hamiltonian to be diagonalizable at all times precludes the application of CD driving in important scenarios where such knowledge is unavailable, e.g., in quantum annealing. However, the development of approximate methods to engineer CD controls has challenged and de facto removed this requirement. Specifically,   the early proposal of using digital methods for quantum simulation to realize CD controls \cite{delCampo12,Saberi14}, in combination with variational methods \cite{Sels17,Claeys19,Bukov21}, has led to a framework for digitized-CD quantum driving for quantum algorithms \cite{Hegade21factorization,Hegade21Portfolio}, that include the use of STA in adiabatic quantum computation  \cite{Hegade21} and quantum optimization  \cite{Chandarana22,Hegade22DCQO,Chandarana22-2}.

The nature of the CD controls remains currently an issue. In a system with many degrees of freedom, finding an efficient prescription to determine the CD fields is generally challenging.
The first works exploring STA by CD driving in many-body quantum systems showed that the CD controls generally involved many-body interactions of arbitrary rank (one-body, two-body, etc.) \cite{delCampo12,Takahashi13,Saberi14,Damski14}. In addition, CD terms are generally spatially nonlocal \cite{Saberi14}. In systems of continuous variables, such as a harmonic oscillator or ultracold gases, CD terms cannot always be realized by applying an external potential \cite{Chen10,Dupays21} but may involve nonconservative momentum-dependent Hamiltonian terms \cite{Jarzynski13,delcampo13cd,Deffner14}. Likewise, in spin systems, CD terms may involve interactions among distant spins \cite{delCampo12,Takahashi13,Saberi14,Damski14}. As a result, one of the pressing problems in the development of STA is to find systematic approaches to tailor CD terms. One option is to find unitarily equivalent Hamiltonians for which STAs can be implemented exactly with experimentally available resources \cite{Ibanez12,delcampo13cd,Deffner14}. Another relies on approximate protocols, determined through variational methods  \cite{Opatrny14,Saberi14,Sels17,Claeys19,Hartmann19,YangPang22,Callum23} or otherwise \cite{Takahashi13,Campbell15,Mukherjee16}. 

Current general approaches to engineering STA by CD driving are blind to any structure or symmetry in the actual dynamics. However, it is known that the presence of dynamical symmetries in a given process can significantly simplify the CD protocols required to control it and render the implementation of STA experimentally realizable. 
In cases where a dynamical symmetry is known, one can identify the CD controls in terms of the elements of a closed Lie algebra \cite{Torrontegui14}. However, the application of this approach has been limited to the restricted set of examples in which dynamical symmetries are known, i.e., few-level systems \cite{MartinezGaraot14} and scale-invariant processes \cite{BeauDelcampo20}.

Further progress calls for novel approaches that systematically unravel and exploit any structure in the dynamics of the process to be controlled. 
This work introduces an approach that achieves this goal by formulating CD driving in Krylov space.
Krylov subspace methods have a long tradition in numerical recipes and can be efficiently implemented using the Lanczos algorithm and its variants \cite{Liesen12}. In time-dependent quantum mechanics, Krylov space describes the minimal subspace in which the dynamics unfolds, greatly easing the computational resources to describe time evolution \cite{Viswanath94}. They are further useful in foundations of quantum physics to characterize operator growth \cite{Parker19,Barbon19,Rabinovici21,Caputa22} and the fundamental speed limits governing it \cite{Hornedal22,Hornedal23}.
Consider the case of the quantum dynamics in the Heisenberg representation. Given an observable of interest $O_0$ and a generator of evolution $H$, the evolution of the observable is set by $O(t)=U^\dag(t) O_0U(t)$ where $U(t)$ is the time-evolution operator. For time-independent Hamiltonians such evolution admits the expansion $O(t)=\sum_{n=0}^\infty (it)^n \mathcal{L}^n O_0/n!$ with the Liouvillian $\mathcal{L}(\cdot)=[H,\cdot]$. The dynamics generates the set of operators $\{\mathcal{L}^n O_0\}_{n=0}^\infty$ that are not orthonormal and generally live in an operator subspace known as the Krylov space. 
The Lanczos algorithm can be used to construct a basis in Krylov space and further provides the Lanczos coefficients that determine the entries of the matrix representation of the Liouvillian in the Krylov basis.

\section{Outline}

In this work, we introduce a formulation of CD driving in Krylov space 
using the celebrated Lanczos algorithm. 
In Sec.~\ref{sec:agp}, we briefly review the key concept of STA.
In the CD driving, for a given time-dependent Hamiltonian, 
the dynamics is assisted by an auxiliary control field known as the CD term. 
The CD Hamiltonian acts as the generator of adiabatic continuation, 
discussed in proofs of the adiabatic theorem, e.g., by Kato~\cite{Kato50} 
and Avron and Elgart~\cite{Avron99}. 
Similarly, it has been discussed in the context of quasiadiabatic continuation by 
Hastings~\cite{Hastings04,Hastings05,Hastings2007Quasi,Hastings2010Locality,
Hastings2015Quantization,DeRoeck15,Bachmann17,Bachmann18}.
Recent literature refers to the CD term as the adiabatic gauge potential (AGP). 
It is further related to the Berry connection, and its norm gives the real part of 
the quantum geometric tensor, i.e., the quantum metric tensor or fidelity susceptibility, 
as discussed in Refs.~\cite{Demirplak08,delCampo12,Funo17,delCampo2018}. 
Thus, the AGP has broad applications beyond quantum control, 
extending to quantum state distinguishability, quantum state geometry, 
adiabatic theorems, critical phenomena, quantum thermodynamics, etc.

Finding the explicit form of the AGP is a fundamental problem 
for practical applications and has been discussed from various viewpoints.
The spectral representation obtained in the original 
studies~\cite{Demirplak03, Demirplak05, Demirplak08, Berry09} has a disadvantage as 
it is generally difficult to obtain the corresponding operator form 
in systems with many degrees of freedom.
In that case, we can start the analysis from the operator equation for the AGP.
The equation is solved approximately using the variational method~\cite{Sels17}.

In the variational method, the validity of the approximation 
strongly depends on the chosen ansatz.
The operator equation is recast into an integral representation.
It gives a nested commutator expansion and 
indicates possible operator forms of the AGP.
Combining the nested commutator expansion with the variational method
offers a systematic method for finding the AGP for complex systems~\cite{Claeys19}.
In most applications,  the expansion is truncated to obtain an approximate result.
It is implied that taking into the infinite-order expansion gives the exact result, 
although rigorous proof has not been shown.
In the present work, we circumvent the need for the variational method 
by providing an exact closed-form expression for the AGP in the Krylov method.

In Sec.~\ref{sec:krylov}, 
we introduce the basic concept of the Krylov subspace method
and develop a general framework for finding the exact AGP from the Krylov expansion.
The Krylov expansion is formulated by defining a proper inner product and 
a Liouvillian superoperator for a target system.
For a given initial seed operator, the Krylov subspace where the dynamics unfolds
is determined from the Krylov algorithm.
We show that a specific choice of the seed operator is useful 
to solve the equation for the AGP.
The AGP is expressed by the Krylov basis and the Lanczos coefficients
obtained from the Krylov algorithm.
We find that the AGP is classified into two categories.
They are characterized by the parity of the number of the Krylov basis.

Comparing the exact form of the AGP 
with the integral representation with the nested commutator expansion
gives a close relation of the AGP to the complexity of the Krylov space.
We discuss that the properties of the AGP can be understood directly from 
the series of the Lanczos coefficients and 
the operator wave functions defined from the general framework of the Krylov method.
We also discuss how the variational method with the nested commutator expansion is justified.

In Sec.~\ref{sec:small}, 
we apply the general framework to various canonical examples, 
including two- and three-level systems, and the harmonic oscillator. 
To be instructive, we demonstrate those well-known examples 
by using several different ways to determine the AGP.

The full potential of the present framework is displayed 
when it is applied to systems with many degrees of freedom.
In Sec.~\ref{sec:many}, 
we treat integrable, nonintegrable, and disordered quantum spin chains.
We first apply the method to a one-dimensional transverse Ising model 
without a longitudinal magnetic field.
The AGP of the system is well known in that case~\cite{delCampo12, Takahashi13,Damski14} 
and we rewrite the result with respect to the Lanczos coefficients.
We find that the quantum phase transition can be identified from 
the Lanczos coefficient series.
When we apply the longitudinal magnetic field, the system becomes nonintegrable and 
the exact solution is not available.
We consider a truncation of the Krylov expansion, and the result is shown to be 
equivalent to that of the variational method. 
In reporting explicit expressions for the AGP in many-body systems, 
our work advances the study of STA beyond the large body of literature 
focused on leading-order truncations of 
the CD term~\cite{Sels17,Hartmann19,Claeys19,Callum23}.  

We also discuss in the same section the one-dimensional isotropic XY model.
We treat several cases where the interaction couplings are uniform or random.
Although the model can be mapped onto a free fermion model, the explicit construction
of the AGP for a given set of coupling constants is a difficult task.
We can formulate the expansion systematically and 
demonstrate the expansion up to a considerably large system size.
We discuss closely how each order of the expansion affects the result.
We also consider the case where the system is equivalent to the integrable system 
described by the Toda equations.
We discuss the implications of the integrability condition on the Krylov expansion. 

The present study is concluded with final remarks in
Sec.~\ref{sec:summary}.

\section{Adiabatic gauge potential and counterdiabatic driving}
\label{sec:agp}

Consider a closed quantum system described by the Hamiltonian operator
$H(\lambda)$ depending on the set of parameters 
$\lambda=(\lambda_1,\lambda_2,\dots)$.
Throughout this paper,
a capital letter denotes an operator or a matrix.
Let $|n(\lambda)\rangle$ represent an eigenstate of the Hamiltonian
with the eigenvalue $\epsilon_n(\lambda)$.
The time-independent Schr\"odinger equation, 
and the equation for adiabatic continuation read, respectively, 
\be
 && H(\lambda)|n(\lambda)\rangle = \epsilon_n(\lambda)|n(\lambda)\rangle, \\
 && A(\lambda)|n(\lambda)\rangle = i\partial_\lambda|n(\lambda)\rangle.
\ee
The phase of the eigenstate is fixed by requiring the relation 
$\langle n(\lambda)|\partial_\lambda n(\lambda)\rangle=0$.
The AGP operator $A=(A_1,A_2,\dots)$
is introduced by differentiating the eigenstate with respect to $\lambda$ and 
enforces adiabatic continuation for all eigenstates; i.e., it is independent of $n$.

One of the prominent applications of the AGP is the CD 
driving~\cite{Demirplak03, Demirplak05, Demirplak08, Berry09, Nakahara22}.
For time-varying parameters $\lambda(t)$, we consider the time evolution 
\be
 i\partial_t|\psi(t)\rangle
 =\{H[\lambda(t)]+H_{\rm CD}(t)\}|\psi(t)\rangle.
 \label{tdseCD}
\ee
Here, the CD term is introduced as 
$H_{\rm CD}(t)=\dot{\lambda}(t)\cdot A[\lambda(t)]$, 
where the overdot denotes the time derivative.
It prevents the nonadiabatic transitions among eigenstates $|n(\lambda)\rangle$, 
which means that the solution of the Schr\"odinger equation (\ref{tdseCD}) 
is exactly given by the adiabatic state of $H$: 
\begin{eqnarray}
 |\psi(t)\rangle
 & =& \sum_n e^{-i\int_0^t ds\,\epsilon_n[\lambda(s)]
 -\int_0^t ds \langle n[\lambda(s)]|\partial_sn[\lambda(s)]\rangle}\nonumber\\
 & &\times |n[\lambda(t)]\rangle
 \langle n[\lambda(0)]|\psi(0)\rangle.
\end{eqnarray}
Even when the implementation of the CD term is challenging,
we can use the CD term to assess
the nonadiabatic effects~\cite{Suzuki20}.

While it is a nontrivial problem to obtain
the explicit form of the AGP for a given Hamiltonian, its matrix elements can be 
formally written in terms of the spectral properties of $H$ as
\be
 \langle m(\lambda)|A(\lambda)|n(\lambda)\rangle
 =i\frac{\langle m(\lambda)|\partial_\lambda H(\lambda)|n(\lambda)\rangle}
 {\epsilon_n(\lambda)-\epsilon_m(\lambda)}(1-\delta_{m,n}).
 \label{Asp}
\ee
The main aim of the present work is to find a systematic way to obtain
the operator form of the AGP.
The AGP satisfies 
\be
 \mathcal{L}_\lambda \left[
 \partial_\lambda H(\lambda)-i\mathcal{L}_\lambda A(\lambda)\right]=0,
 \label{eqfora}
\ee
where $\mathcal{L}_\lambda(\cdot)=[H(\lambda),\cdot]$.
This relation was used to find an approximate AGP by variational 
methods~\cite{Sels17, Claeys19, Hatomura21}.
We exploit this relation to obtain the exact form of the AGP.

As an alternative useful relation, one can invoke the integral 
representation introduced by Hastings in the context of quasiadiabatic 
continuation~\cite{Hastings04,Hastings05,Hastings2010Locality}: 
\be
 A(\lambda) &=& -\frac{1}{2}\lim_{\eta\to 0}
 \int_{-\infty}^\infty ds\,{\rm sgn}\,(s)e^{-\eta|s|}
 \no\\ &&\times
 e^{iH(\lambda)s}\partial_\lambda H(\lambda)e^{-iH(\lambda)s}. \label{int}
\ee
The integrand is proportional to the operator $\partial_\lambda H(\lambda)$ 
conjugated by a unitary.
Using the unitary operation is represented by $\mathcal{L}_\lambda$, 
we can perform the integration over $s$ to write 
\be
 A(\lambda) = -\frac{1}{2}\lim_{\eta\to 0}\left(
 \frac{1}{\eta-i\mathcal{L}_\lambda}-\frac{1}{\eta+i\mathcal{L}_\lambda}\right)
 \partial_\lambda H(\lambda).
\ee
This formal expression motivates us to use the 
expansion~\cite{Claeys19,Pandey20,WurtzLove22}
\be
 A(\lambda)=i\sum_k \alpha_k^{\rm nc}(\lambda)
 \mathcal{L}_\lambda^{2k-1}\partial_\lambda H(\lambda). \label{var}
\ee
The construction of the AGP in Krylov space that we present in the following follows 
solely from using the expansion (\ref{var}) in combination with Eq. (\ref{eqfora}). 
Its importance relies on the fact that it removes the need for 
the spectral properties of $H(\lambda)$  in determining the CD term and shows that 
the operators in the AGP are generated from the nested commutators
$\mathcal{L}^{2k-1}\partial H$ at odd orders.
We note that the variable $s$ in the integral representation
represents a fictitious time. The unitary 
$e^{-iH(\lambda)s}$ is interpreted as the time evolution operator
in the fictitious time with no need for the time-ordered product, 
as $H(\lambda)$ is independent of $s$.
When we keep all possible operators generated from the nested commutators,
the exact AGP can be obtained by solving the equation
for $\alpha_k^{\rm nc}$ from Eq.~(\ref{eqfora}).
Practically,  a truncation of the operator series
yields an approximate AGP.
The infinite series by nested commutators produces the same type of operators 
many times, and it is not clear how many terms  should be kept  
to obtain a required accuracy.
To treat the AGP systematically,  
we rearrange the expansion in Eq.~(\ref{var}) and represent
the AGP in a finite series 
by using a set of orthonormal Krylov basis elements.

\section{Krylov expansion}
\label{sec:krylov}

\subsection{Inner product, basis operators, and vector representation of operators} 

In the Krylov method~\cite{Viswanath94}, 
we use a set of operators satisfying an orthonormal relation.
To define the orthonormality of operators,
we first introduce the inner product for an arbitrary pair of operators
$X$ and $Y$ as 
\be
 (X,Y)=\frac{1}{2}{\rm Tr}[\rho(H) (X^\dag Y+YX^\dag)].
 \label{innerprod}
\ee
The operators are not necessarily Hermitian. 
In addition, the measure  $\rho(H)$ is a positive-definite Hermitian operator 
but not necessarily normalized.
We note that the present method is applicable 
even when the Hilbert space dimension 
is infinite and the energy spectrum is continuous, 
provided that $\rho(H)$ is chosen appropriately.
We see in the following that the result of the AGP is independent
of the choice of $\rho(H)$.
What is important is that $\rho(H)$ commutes with $H$.
We have 
\be
 && (X,\mathcal{L} X)=0, \label{diag} \\
 && (X,\mathcal{L} Y)^*=(Y,\mathcal{L} X), 
\ee
for Hermitian operators $X$ and $Y$.

To find an explicit representation of the superoperator $\mathcal{L}$,
we introduce a set of basis operators 
$X=(X_1,X_2,\dots)$, that are Hermitian and orthonormal with each other:
\be
 (X_\mu,X_\nu)=\delta_{\mu,\nu}.
\ee
The number of operators is not specified here and
is discussed in the following after clarifying the aim of the analysis.
Generally, for a given quantum system, it is equal to or smaller than
the square of the dimension of the Hilbert space.

One of the aims of introducing the basis operators is
that the superoperator $\mathcal{L}$ can be represented
by an antisymmetric Hermitian matrix $L$. 
It has elements
\be
 L_{\mu\nu}=(X_\mu,\mathcal{L} X_\nu)
\ee
and satisfies $L^\dag=L$ and $L^{\rm T}=-L$.
The diagonal components are equal to zero and 
each of the off-diagonal components is purely imaginary.
Corresponding to the matrix representation of superoperator, 
a vector represents an operator.
We write the Hamiltonian 
\be
 H=h\cdot X &\Leftrightarrow& |H\rangle= (h_1, h_2,\dots )^{\rm T}
\ee
and the AGP 
\be
 A=a\cdot X &\Leftrightarrow& |A\rangle= (a_1, a_2,\dots )^{\rm T}.
\ee
Then, we obtain a vector representation of Eq.~(\ref{eqfora}) as
\be
 L\left(|\partial_\lambda H\rangle -iL|A\rangle\right)=0. \label{eqfora2}
\ee
It is not a difficult problem to obtain the formal solution
of this equation by using the spectral representation of $L$.
However, $L$ is generally a matrix of large size and
the diagonalization is much more difficult 
than that of the Hamiltonian $H$.
We resolve this problem in the following by introducing an algorithmic method.

\subsection{Lanczos algorithm and Krylov basis}

Equation (\ref{eqfora2}) implies that the AGP $|A\rangle$ is constructed from 
a linear combination of $L^n|\partial H\rangle$ with $n=1,2,\dots$.
We prepare the normalized vector $|\theta_0\rangle$ from the relation 
\be
 b_0|\theta_0\rangle = |\partial H\rangle. \label{theta0}
\ee 
The coefficient $b_0$ represents the normalization factor and is
written as 
\be
 b_0^2 = \langle\partial H|\partial H\rangle = (\partial H,\partial H).
\ee
The zeroth-order normalized vector $|\theta_0\rangle$ and the coefficient $b_0$ are defined
for each component of $\lambda$.
The same applies to the quantities introduced in the following.
We abbreviate the component index to simplify the notation.
Then, the new normalized basis $|\theta_1\rangle$ is defined from 
\be
 b_1|\theta_1\rangle = L|\theta_0\rangle.
\ee
By construction, $|\theta_1\rangle$ is orthogonal to $|\theta_0\rangle$.  
We repeat the same procedure by using the relation 
\be
 b_n|\theta_n\rangle = L|\theta_{n-1}\rangle -b_{n-1}|\theta_{n-2}\rangle,
\ee
with $n=2,3,\dots$.
The positive coefficient $b_n$ is chosen
so that $|\theta_n\rangle$ is normalized.
Thus, the introduced vectors satisfy the orthonormal relation
$\langle\theta_m|\theta_n\rangle=\delta_{m,n}$.
When the dimension of the Hilbert space is finite, 
the number of basis elements must be finite, which means that
there exists an integer $d$ satisfying 
$L|\theta_{d-1}\rangle-b_{d-1}|\theta_{d-2}\rangle=0$.
The number of the basis vectors is given by $d$, 
which we refer to as the Krylov dimension.

This way of constructing a basis set is nothing but 
the Lanczos algorithm since the matrix $L$ is brought to a tridiagonal form $T$, 
satisfying $L=VTV^\dag$, where 
$V=(|\theta_0\rangle\ |\theta_1\rangle\ \dots\ |\theta_{d-1}\rangle)$ and 
\be
 T = \left(\begin{array}{cccccc}
 0 & b_1 & 0 &&& \\ b_1 & 0 & b_2 &&& \\ 0 & b_2 & 0 & && \\
 &&& \ddots && \\ &&& & 0 & b_{d-1} \\ &&&& b_{d-1} & 0
 \end{array}\right). \label{T}
\ee
We can also write 
\be
 L=\sum_{n=1}^{d-1} b_{n}\left(
 |\theta_{n}\rangle\langle\theta_{n-1}|+|\theta_{n-1}\rangle\langle\theta_{n}|
 \right).
\ee
Generally, for a given matrix $L$ and an initial basis element $|\theta_0\rangle$, 
we can render the matrix in tridiagonal form algorithmically.
We find in the following that
the present choice of the initial basis in Eq.~(\ref{theta0}) is
convenient to solve Eq.~(\ref{eqfora2}).

The introduction of the orthonormal basis vectors corresponds to
that of the orthonormal basis operators 
$|O_n\rangle=|\theta_n\rangle$.
In the original representation, 
\be
 O_n =\theta_n\cdot X, 
\ee
with $n=0,1,2,\dots,d-1$.
They are generated by the procedure 
\be
 \begin{array}{ll}
 b_0O_0 = \partial H, & \\
 b_1O_1 = \mathcal{L}O_0, & \\
 b_nO_n = \mathcal{L}O_{n-1}-b_{n-1}O_{n-2} & (n=2,3,\dots,d-1)
 \end{array}
 \label{krylovbasis}
\ee
and satisfy $(O_m,O_n)=\langle\theta_m|\theta_n\rangle=\delta_{m,n}$.
This set of operators represents the Krylov basis. 
In the present choice of $O_0$,
the operators of even order $O_{2k}$ ($k=0,1,2,\dots$) are Hermitian, and
those of odd order $O_{2k-1}$ ($k=1,2,\dots$) are anti-Hermitian.

We note that the introduction of the basis operators $X$
is not necessary, since we can construct the Krylov basis directly
from Eq.~(\ref{krylovbasis}).
The introduction of the basis operators makes it clear that
the introduction of the Krylov basis is equivalent
to the Lanczos algorithm.
The following examples illustrate that the two options can prove convenient.

The advantage of the basis operator representation of $\mathcal{L}$ by $L$ 
is that we do not need to calculate the nested commutators
$\mathcal{L}^{n}\partial H$ once we can construct a single matrix $L$.
We also see that the number of the basis operators $X$ is not necessarily
equal to the square of the dimension of the Hilbert space $d_H$.
For the present purpose, we need operators in $\mathcal{L}^{n}\partial H$ and 
the dimension of $L$, denoted by $d_L$, satisfies 
\be
 d\le d_L\le d_H^2.
\ee
Thus, the Krylov dimension $d$ is defined by the minimum number of the basis elements.
When the matrix $L$ is block diagonalized, we may treat only
the block in which the operators in $\partial H$ are included.
A good choice of the basis reduces the computational cost. 
It is known that the general upper limit of the Krylov dimension is 
given by the relation $d\le d_H^2-d_H+1$~\cite{Rabinovici21}.

Generally, the Krylov method is useful
when we treat the Heisenberg representation of a normalized operator $O_0$,
$O(s)=e^{iHs} O_0e^{-iHs}$~\cite{Viswanath94, Parker19, Barbon19, 
Rabinovici21, Caputa22}.
We can represent the operator by a finite series as 
\be
 O(s)=\sum_{n=0}^{d-1}i^n\varphi_n(s)O_n,
\ee
where $\varphi_n(s)$ is known as the operator wave function.
The time dependence of $O(s)$  can be conveniently studied
by using the operator wave function, a feature we next apply to 
the computation of the AGP from the integral representation in  Eq. (\ref{int}). 

\subsection{Adiabatic gauge potential}

We are now in a position to solve Eq.~(\ref{eqfora}),
or the equivalent Eq.~(\ref{eqfora2}), by using the Krylov basis.
We use 
\be
 A=ib_0\sum_{k=1}^{d_A} \alpha_kO_{2k-1} &\Leftrightarrow&
 |A\rangle = ib_0\sum_{k=1}^{d_A} \alpha_k|\theta_{2k-1}\rangle
 \no\\ \label{agpk}
\ee
and solve the equation for $\{\alpha_k\}_{k=1}^{d_A}$.
It is important to notice that $A$ includes the Krylov basis
at odd order, $O_{2k-1}$.
This property is a direct consequence of the representation
in Eq.~(\ref{var}).
The number of the operators is denoted by $d_A$ and is related
to the Krylov dimension $d$ as 
\be
 d_A= \left\lfloor{\frac{d}{2}}\right\rfloor
 = \left\{\begin{array}{ll}
 d/2 & \mbox{for even $d$,}\\
 (d-1)/2 & \mbox{for odd $d$.} \end{array}\right.
\ee

We first consider the case of even $d$.
In this case, one finds 
\be
 && |\partial_\lambda H\rangle-iL|A\rangle
 = b_0(1+\alpha_1b_1)|\theta_0\rangle
 \no\\ &&
 +b_0\sum_{k=1}^{d_A-1}\left(\alpha_k b_{2k}+\alpha_{k+1} b_{2k+1}\right)
 |\theta_{2k}\rangle. \label{aeven}
\ee
Setting each side of this equation to zero yields
\be
 && \alpha_1 = -\frac{1}{b_1}, \label{alphaevend1} \\
 && \alpha_{k+1}= -\frac{b_{2k}}{b_{2k+1}}\alpha_{k}, \label{alphaevend2}
\ee
where $k=1,2,\dots,d_A-1$.
That is, we can find the AGP satisfying the relation 
$|\partial H\rangle -iL|A\rangle=0$, which is a sufficient condition
of Eq.~(\ref{eqfora2}).
We also see that the relation 
$|\partial H\rangle -iL|A\rangle=0$ represents the equation for
a dynamical invariant, when 
the eigenvalues of the Hamiltonian, $\epsilon_n[\lambda(t)]$,
are time independent~\cite{Lewis69,Takahashi22}.
In this case, diagonal components of $\partial_\lambda H(\lambda)$ 
in the eigenstate basis are equal to zero, i.e., 
$\langle n(\lambda)|\partial_\lambda H(\lambda)|n(\lambda)\rangle=0$.

Next, we consider the case of odd $d$.
In this case, an additional term appears in Eq.~(\ref{aeven})
and no solution exists for $|\partial H\rangle -iL|A\rangle=0$.
We examine $L^2|A\rangle = -iL|\partial H\rangle$ to find the expression 
\begin{widetext}
\be
 \left(\begin{array}{cccccc}
 b_1^2+b_2^2   & b_2b_3 & 0 &&&\\
 b_2b_3 & b_3^2+b_4^2 & b_4b_5 &&&\\
 0 & b_4b_5 & b_5^2+b_6^2 & &&\\
 &&&\ddots && \\
 &&& & b_{d-4}^2+b_{d-3}^2 & b_{d-3}b_{d-2} \\
 &&&& b_{d-3}b_{d-2} & b_{d-2}^2+b_{d-1}^2
 \end{array}\right)
 \left(\begin{array}{c}
 \alpha_1 \\ \alpha_2 \\ \vdots \\ \alpha_{d_{A}}
 \end{array}\right)
 =\left(\begin{array}{c}
 -b_1 \\ 0 \\ \vdots \\ 0
 \end{array}\right). \label{alphaoddd}
\ee
\end{widetext}
Inverting the matrix in this expression, 
we can obtain the explicit form of the AGP.
In the following, we solve this equation by using a different approach
which proves illuminating. 

\begin{widetext}
\begin{figure*}[t]
\includegraphics[width=2\columnwidth]{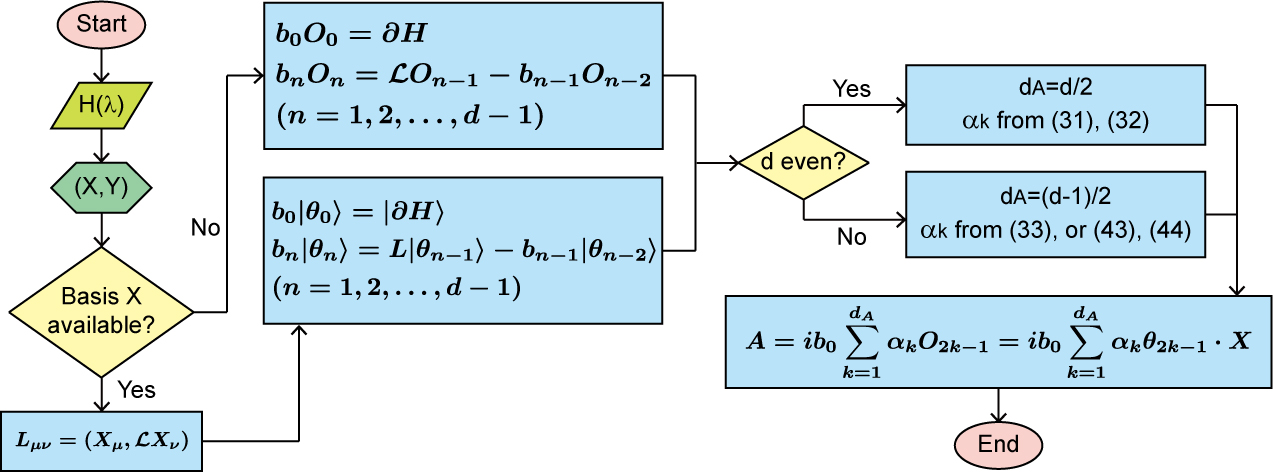}
\caption{
The flowchart of the Krylov algorithm 
to obtain the AGP $A(\lambda)$ for a given Hamiltonian $H(\lambda)$.
}
\label{fig01}
\end{figure*}
\end{widetext}

We conclude this part by stating that 
the AGP can be constructed systematically by using the Krylov basis.
The AGP is represented by an expansion of the Krylov basis, 
and the coefficient of each term is obtained as a function of
$b_0$, the scale of $\partial H$, and 
the set of Lanczos coefficients $\{b_n\}_{n=1}^{d-1}$.
When $d$ is even, the instantaneous eigenvalues of the Hamiltonian 
must be time independent.
Conversely, the Krylov dimension $d$ is even (odd)
when the eigenvalues of the Hamiltonian are time independent (dependent).
The flowchart of the algorithm is presented in Fig.~\ref{fig01}.

Equation~(\ref{agpk}) is compared with Eq.~(\ref{var}).
The former is expanded by orthonormal operators and 
the total number of series elements is finite, 
if the resulting AGP is given by a finite number of operators.
The expansion is also applicable to systems with a continuous spectrum.
Thus, the Krylov method offers a general systematic method
for constructing the AGP.

\subsection{Operator wave function and adiabatic gauge potential}
\label{owf}

The AGP is closely related to the operator wave function $\varphi_n(\lambda,s)$
defined from the Heisenberg representation 
\be
 e^{iH(\lambda)s}O_0(\lambda)e^{-iH(\lambda)s}
 =\sum_{n=0}^{d-1}i^n\varphi_n(\lambda,s)O_n(\lambda),
\ee
where the initial condition is chosen as
$b_0(\lambda)O_0(\lambda)=\partial_\lambda H(\lambda)$.
Substituting this representation into Eq.~(\ref{int}), we obtain 
\be
 \frac{1}{2}\lim_{\eta\to 0}\int_{-\infty}^\infty ds\,
 {\rm sgn}\,(s)e^{-\eta|s|}\varphi_{2k}(\lambda,s)=0,
\ee
for $k=0,1,\dots,d_A$, and 
\be
 \frac{1}{2}\lim_{\eta\to 0}\int_{-\infty}^\infty ds\,
 {\rm sgn}\,(s)e^{-\eta|s|}\varphi_{2k-1}(\lambda,s)=(-1)^{k}\alpha_k(\lambda),
 \no\\ \label{phia}
\ee
for $k=1,2,\dots,d_A$.
This relation between $\varphi_n(\lambda,s)$ and $\alpha_k(\lambda)$
shows that the latter is obtained from the Laplace transform of the former.
The behavior of the operator wave function has been studied in the context
of the Krylov complexity, and we can exploit the properties obtained 
in that context~\cite{Viswanath94, Parker19, Barbon19, Rabinovici21, Hornedal22,Caputa22}.
For example, the operator wave function
$|\varphi(\lambda,s)\rangle= (\varphi_0,\varphi_1,\dots,\varphi_{d-1})^{\rm T}$
satisfies the differential equation 
$\partial_s|\varphi(\lambda,s)\rangle=B(\lambda)|\varphi(\lambda,s)\rangle$
with
\be
 B=\left(\begin{array}{cccccc}
 0 & -b_1 & 0 &&&\\ b_1 & 0 & -b_2 &&&\\ 0 & b_2 & 0 & &&\\
 &&&\ddots && \\ &&& & 0 & -b_{d-1} \\ &&&& b_{d-1} & 0
 \end{array}\right)
 \label{b}
\ee
and the initial condition $|\varphi(\lambda,0)\rangle=(1,0,0,\dots)^{\rm T}$.
Here, $iB$ is related to the matrix $T$ in Eq.~(\ref{T})
under a unitary transformation.
Since the equation for $|\varphi(\lambda,s)\rangle$ is interpreted as
a Schr\"odinger equation with a
Hamiltonian $iB(\lambda)$ independent of the fictitious time $s$, 
the solution is obtained by solving the eigenvalue problem 
$iB(\lambda)|\omega_n(\lambda)\rangle=\omega_n(\lambda)|\omega_n(\lambda)\rangle$.
We can write 
\be
 |\varphi(\lambda,s)\rangle=\sum_{n=0}^{d-1}e^{-i\omega_n(\lambda)s}
 |\omega_n(\lambda)\rangle\langle\omega_n(\lambda)|\varphi(\lambda,0)\rangle.
\ee
The form of the Hermitian matrix $iB$ indicates
that the eigenvalues come in pairs $\pm \omega_n$ where $\omega_n\ne 0$,
and the zero-eigenvalue state exists only
when the size of the matrix $d$ is odd.
We refer to the details on the pairing of eigenstates in Appendix~\ref{deriveAA}.
Here, we look at only the zero-eigenvalue state $|\phi(\lambda)\rangle$
satisfying $B(\lambda)|\phi(\lambda)\rangle=0$ for odd $d$.
We can solve the eigenvalue equation  to obtain the normalized
solution 
\be
 |\phi\rangle=\frac{1}{\sqrt{1+\left(\frac{b_1}{b_2}\right)^2+\cdots}}
 \left(\begin{array}{c}
 1 \\ 0 \\ \frac{b_1}{b_2} \\ 0 \\ \frac{b_3b_1}{b_4b_2} \\ \vdots \\
 \frac{b_{d-2}b_{d-4}\dots b_1}{b_{d-1}b_{d-3}\dots b_2}
 \end{array}\right).
 \label{phi}
\ee
Since the matrices $L$ and $B$ are constructed from
the commutator $\mathcal{L}(\cdot)=[H,(\cdot)]$,
the eigenvalues are related to
the energy eigenvalue difference $\epsilon_m-\epsilon_n$.
The zero-eigenvalue state of $M$ implies the existence of
the diagonal components 
\be
 \partial H-i\mathcal{L}A= \sum_n\partial\epsilon_n |n\rangle\langle n|.
\ee
The contribution on the right-hand side 
is absent for even $d$ with $\partial\epsilon_n=0$.

We can also use the equation for $|\varphi(\lambda,s)\rangle$
to obtain the explicit form of $\alpha_k$.
The $2k$th component of the equation is given by  
$\partial_s \varphi_{2k}=b_{2k}\varphi_{2k-1}-b_{2k+1}\varphi_{2k+1}$.
Using the integral representation in Eq.~(\ref{int}), we obtain 
\be
 \lim_{\eta\to 0}\int_{0}^\infty ds\,e^{-\eta s}\partial_s\varphi_{2k}
 =(-1)^{k}\left(b_{2k}\alpha_k+b_{2k+1}\alpha_{k+1} \right).
 \no\\
\ee
The left-hand side is calculated by using the integration by parts
to give 
\be
 \lim_{\eta\to 0}\int_{0}^\infty ds\,e^{-\eta s}\partial_s\varphi_{2k}
 =-\delta_{k,0}+\phi_{2k}\phi_0,
\ee
where the second term exists only for odd $d$ and we write
$|\phi\rangle=(\phi_0,\phi_1,\dots)^{\rm T}$.
In the odd-$d$ case, we obtain 
\be
 && \alpha_1=-\frac{1}{b_1}+\frac{\phi_0^2}{b_1}, \label{alphaoddd1}\\
 && \alpha_{k+1}=-\frac{b_{2k}}{b_{2k+1}}\alpha_k
 +\frac{(-1)^{k}\phi_{2k}\phi_0}{b_{2k+1}},
 \label{alphaoddd2}
\ee
with $k=1,2,\dots,d_A-1$.
It is not a difficult task to confirm that this relation is consistent
with Eq.~(\ref{alphaoddd}).

The use of the operator wave function also allows us to obtain 
\be
 (A,A)=b_0^2\sum_{k=1}^{d_A}\alpha_k^2 = b_0^2\langle 0|(QiBQ)^{-2}|0\rangle,
 \label{AA}
\ee
where $|0\rangle=(1,0,\dots,0)^{\rm T}$ and 
$Q=1-|\phi\rangle\langle\phi|$ represents the projection operator
onto the nonzero-eigenvalue states.
We show the derivation in Appendix~\ref{deriveAA}.
This representation is useful when we evaluate the norm of the AGP.

It is instructive to compare the present result for the odd-dimension case
to that for the even-dimension case.
Equations~(\ref{alphaevend1}) and (\ref{alphaevend2}) show that, 
when the Krylov dimension is even, each order is calculated without using 
the higher-order contributions. 
This property is practically useful for systems with many degrees of freedom.
As we discuss in the next sections, we frequently 
consider the truncation of the series expansion as an approximation.
By contrast, for an odd Krylov dimension, all the Lanczos coefficients  
are required to construct each term of the AGP, as we see in 
Eqs.~(\ref{alphaoddd1}) and (\ref{alphaoddd2}) and 
the zero mode $|\phi\rangle$ in Eq.~(\ref{phi}).
However, each component of $|\phi\rangle$ takes a small value
and could be negligible for large systems.

\begin{figure}[t]
\centering\includegraphics[width=1.\columnwidth]{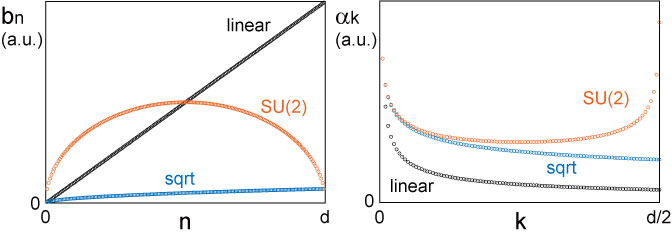}
\caption{
The Lanczos coefficients $b_n$ (left) 
and the coefficients of the AGP $\alpha_k$ (right).
We show the cases $b_n\propto n$ (linear), 
$b_n\propto \sqrt{n}$ (sqrt), and $b_n\propto \sqrt{n(d-n)}$ (SU(2)).
}
\label{fig02}
\end{figure}

We can estimate a contribution from each term
of the expansion in Eq.~(\ref{agpk}) by the Lanczos coefficient.
When $b_n$ is an increasing function with respect to $n$,
the corresponding $\alpha_k$ is a decreasing function.
The typical global behavior of the Lanczos coefficients has been discussed 
in many-body systems.
It was found that $b_n\propto n$ for chaotic systems 
and leads to a maximal pace of operator growth. 
Likewise, $b_n\propto \sqrt{n}$ for integrable systems, and 
$b_n\approx {\rm const}$ for noninteracting systems \cite{Parker19}.
In Fig.~\ref{fig02}, we show the  behavior of $\alpha_k$ 
in the case of a linear and square-root growth of $b_n$.
The constant case is found in the examples in Sec.~\ref{sec:many}.
In the figure, we also show a special case $b_n\propto\sqrt{n(d-n)}$  
where the operators defined from the Krylov complexity theory 
form a SU(2) algebra~\cite{Parker19,Barbon19,Rabinovici21,Caputa22,Hornedal22,Hornedal23}.
We also note that the series of  Lanczos coefficients typically shows an oscillating
behavior, as shown in the  examples below.
Given that the coefficients $\alpha_k$ in the AGP expansion involve 
the ratio $b_{2k}/b_{2k+1}$ as in Eq. (\ref{alphaoddd2}), 
a regular oscillation series of $b_k$ leads to 
a decreasing series on $\alpha_k$. 
These observations indicate that the property of the CD term is closely related
to that of the operator growth in the Krylov subspace.

\subsection{Classification of basis operators}

It is instructive to notice that the AGP consists
of the nested commutators at odd orders.
When the original Hamiltonian is real symmetric, 
the nested commutators 
at even orders $\mathcal{L}^{2k}\partial H$ are real symmetric and 
those at odd orders $\mathcal{L}^{2k-1}\partial H$
involve the imaginary unit.
This means that the basis operators are classified into two parts: 
\be
 X\to (X;Y)=(X_1,X_2,\dots, X_{d_X};Y_1,Y_2,\dots,Y_{d_Y}).
\ee
$X$ represents basis at even orders and $Y$ at odd orders.
These Hermitian operators satisfy $\mathcal{L}X\in Y$ and $\mathcal{L}Y\in X$.
Accordingly, the matrix $L$, the basis operator representation of $\mathcal{L}$,
takes the form 
\be
 L = \left(\begin{array}{cc}
 0 & M \\ M^\dag & 0 \end{array}\right), \label{LM}
\ee
where $M_{\mu\nu}=(X_\mu,\mathcal{L}Y_\nu)$.
We note that $M_{\nu\mu}^*=(Y_\mu,\mathcal{L}X_\nu)$.
The size of the matrix $M$ is determined by the numbers of
the basis operators $d_X$ and $d_Y=d_L-d_X$.
$M$ is generally a rectangular matrix, and
the Lanczos algorithm is applied for even $d=2d_A$ as 
\begin{widetext}
\be
 M = \left(\begin{array}{cccc}
 |\theta_0\rangle & |\theta_2\rangle & \dots & |\theta_{d-2}\rangle 
 \end{array}\right)
 \left(\begin{array}{cccccc}
 b_1 & 0 & 0 & &&\\ b_2 & b_3 & 0 &&& \\ 0 & b_4 & b_5 &&& \\
 &&&\ddots&& \\ &&& & b_{d-3} & 0 \\ &&& & b_{d-2} & b_{d-1}
 \end{array}\right)
 \left(\begin{array}{c}
 \langle\theta_1| \\ \langle\theta_3| \\ \vdots \\ \langle\theta_{d-1}|
 \end{array}\right),
\ee
where each of $\{|\theta_{2k}\rangle\}_{k=0}^{d_A-1}$ has $d_X$ components, 
each of $\{|\theta_{2k-1}\rangle\}_{k=1}^{d_A}$ has $d_Y$ components,
and the size of the lower triangular matrix on the right-hand side
is $d_A\times d_A$.
Since the numbers of the basis operators must be large enough to
span the operator space, we find $d_X\ge d_A$ and $d_Y\ge d_A$.
In the case of odd $d=2d_A+1$, $M$ is decomposed as 
\be
 M = \left(\begin{array}{cccc}
 |\theta_0\rangle & |\theta_2\rangle & \dots & |\theta_{d-1}\rangle 
 \end{array}\right)
 \left(\begin{array}{cccccc}
 b_1 & 0 & 0 &&& \\ b_2 & b_3 & 0 &&& \\ 0 & b_4 & b_5 &&& \\ &&&\ddots&& \\
 &&&& b_{d-4} & 0 \\ &&&& b_{d-3} & b_{d-2} \\ &&&& 0 & b_{d-1} 
 \end{array}\right)
 \left(\begin{array}{c}
 \langle\theta_1| \\ \langle\theta_3| \\ \vdots \\ \langle\theta_{d-2}|
 \end{array}\right),
\ee
\end{widetext}
where each of $\{|\theta_{2k}\rangle\}_{k=0}^{d_A}$ has $d_X$ components, 
each of $\{|\theta_{2k-1}\rangle\}_{k=1}^{d_A}$ has $d_Y$ components,
and the size of the matrix on the right-hand side is $(d_A+1)\times d_A$.
We also find $d_X\ge d_A+1$ and $d_Y\ge d_A$.
In this case, the minimum number of $d_X$ is larger than that of $d_Y$.

\subsection{Relation to the variational method}

The orthonormal relation of operators 
is useful to understand the relation between the present method
and the variational method~\cite{Sels17}.
In the variational method, the AGP is obtained by minimizing the cost function 
\be
 G[A]={\rm Tr}[\left(\partial H-i{\cal L}A\right)^2],
\ee
for a given operator ansatz of $A$ with undetermined coefficients.
In our notation, this can be written as 
\be
 G[A]=\left(\langle\partial H|+i\langle A|L\right)
 \left(|\partial H\rangle -iL|A\rangle\right),
\ee
with $\rho(H)=1$.
One of the systematic methods for obtaining the AGP
is to use the nested commutator series in Eq.~(\ref{var})
and carrying out the minimization procedure  
with respect to the coefficients $\alpha_k^{\rm nc}(\lambda)$~\cite{Claeys19}.
Practically, the number of series elements in Eq.~(\ref{var}) is restricted to
a finite value, and the approximate AGP is obtained from the minimization.

It is not obvious that the variational method can give the exact AGP
even when all of the possible operators are incorporated 
in the trial form of the AGP by nested commutators.
When the AGP satisfies $|\partial H\rangle-iL|A\rangle=0$,
which is a sufficient condition of Eq.~(\ref{eqfora2}), 
we have discussed that the Krylov dimension is even and
that the matrix $L$ as well as $B$ are invertible.
Then, the minimization procedure gives the exact AGP
$|A\rangle=-iL^{-1}|\partial H\rangle$.

On the other hand, when the Krylov dimension is odd, $L$ is not invertible
and special care is required for the zero-eigenvalue state.
The zero-eigenvalue state of $L$ denoted by $|\phi_L\rangle$
is obtained in the same way as 
that of $B$, Eq.~(\ref{phi}), and is written by  
a linear combination of even basis $\{|\theta_{2k}\rangle\}_{k=0}^{(d-1)/2}$.
It is orthogonal to the AGP as $\langle\phi_L|A\rangle=0$, 
and the cost function is decomposed as
\be
 G[A]&=&\langle\partial H|P|\partial H\rangle \no\\
 &&+\left(\langle\partial H|Q+i\langle A|L\right)
 \left(Q|\partial H\rangle -iL|A\rangle\right),
\ee
where $P=|\phi_L\rangle\langle \phi_L|$ and $Q=1-P$ are projection operators.
The first term does not affect the variational procedure, 
and the minimization of the second term gives 
\be
 |A\rangle = -i(QLQ)^{-1}Q|\partial H\rangle.
\ee
This is the solution of
$L\left(|\partial H\rangle -iL|A\rangle\right)=0$,
which means that the variational method gives the exact AGP.

We note that the trial form of the AGP must include all
possible operators from the nested commutators at odd order
to find the exact AGP from the variational method.
When we consider a restricted number of operators, the minimization
gives an approximate AGP.
This procedure is essentially equivalent to considering a
restricted number of basis operators for the Krylov expansion.
However, as we explicitly show in the following examples, 
the variational procedure does not necessarily lead to 
the result from the Krylov expansion.
This is because the coefficients of the AGP in the variational method 
are optimized in the truncated space.

A significant fact is that we can find the exact AGP 
associated with a generalized cost function of the form
\be
 G[A]={\rm Tr}[\rho(H)\left(\partial H-i{\cal L}A\right)^2].
\ee
This form is useful when the dimension of the Hilbert space is infinite
and when the spectrum is continuous.
Although the exact AGP must be independent of $\rho(H)$, 
the approximate AGP is generally dependent on it.
In the variational method, we usually set a constant $\rho(H)$.
It may be possible to use a different $\rho(H)$ for the variational calculation.
However, even when all possible operators are incorporated, 
it is not evident that the variational method gives the exact result, 
which is independent on the choice of $\rho(H)$. 
The Krylov method states the requirements for $\rho(H)$ explicitly and 
clarifies that the result is independent on that choice.

\section{Applications to small systems}
\label{sec:small}

In the construction of STA, one is interested 
in the time dependence of the Hamiltonian.
We set $\lambda(t)=t$ and identify $\lambda$ as time $t$.
Then, the AGP is equivalent to the CD term.
In the applications discussed below, 
we write the Hamiltonian as $H(t)$ 
and use the CD term $H_{\rm CD}(t)$ instead of the AGP $A(\lambda)$.
For the small systems discussed in the present section,
it is not a difficult task to calculate the CD term explicitly.
We study how the CD term is obtained by the Krylov method 
in typical small systems.

\subsection{Two-level system}

The study of STA by CD driving in the canonical two-level
system~\cite{Demirplak03, Demirplak05, Demirplak08, Berry09}
was soon followed by its experimental demonstration~\cite{Bason12, Zhang13, Du16},
often in a rotating frame, i.e., making use of a unitarily equivalent
CD Hamiltonian.
To illustrate the engineering of STA in Krylov space,
consider the two-level Hamiltonian
\be
 H(t)=\frac{1}{2}h(t)\bm{n}(t)\cdot\bm{\Sigma},
\ee
in terms of the positive scalar $h$, the unit vector $\bm{n}=(n_1,n_2,n_3)$,
and the vector $\bm{\Sigma}=(X,Y,Z)$ with Pauli operators as entries.
In this case,
we have essentially no other choices than to set the basis operators
as $(X,Y,Z)$.
We choose $\rho(H)=1/2$ for the inner product.
We assume that $\bm{n}(t)$ depends on $t$; otherwise, the CD term trivially gives zero.
However, the explicit parameter dependence of $h(t)$ is not necessary, as 
$h$ determines only the overall scale of the Hamiltonian and 
the resulting CD term is independent of $h$.

It is a simple task to calculate the $L$ matrix explicitly.
We have 
\be
 L = ih\left(\begin{array}{ccc}
 0 & -n_3 & n_2 \\ n_3 & 0 & -n_1 \\ -n_2 & n_1 & 0 
 \end{array}\right).
\ee
Then, we set the initial basis vector 
\be
 b_0|\theta_0\rangle
 = \frac{\dot{h}}{2}\left(\begin{array}{c}
 n_1 \\ n_2 \\ n_3 
 \end{array}\right)
 +\frac{h}{2}\left(\begin{array}{c}
 \dot{n}_1 \\ \dot{n}_2 \\ \dot{n}_3 
 \end{array}\right)
\ee
to generate the Krylov basis
\be
 && O_0 = \frac{1}{\sqrt{2}}\frac{\dot{h}\bm{n}+h\dot{\bm{n}}}
 {\sqrt{\dot{h}^2+h^2\dot{\bm{n}}^2}}\cdot\bm{\Sigma}, \\
 && O_1 = \frac{i}{\sqrt{2}}\frac{\bm{n}\times\dot{\bm{n}}}
 {|\dot{\bm{n}}|}\cdot\bm{\Sigma}, \\
 && O_2 = -\frac{1}{\sqrt{2}}\frac{\dot{h}}{|\dot{h}|}
 \frac{1}{\sqrt{\dot{h}^2+h^2\dot{\bm{n}}^2}}
 \left(h|\dot{\bm{n}}|\bm{n}-\frac{\dot{h}\dot{\bm{n}}}{|\dot{\bm{n}}|}
 \right)\cdot\bm{\Sigma} 
 \no\\
\ee
and the Lanczos coefficients 
\be
 && b_1 = \frac{h^2|\dot{\bm{n}}|}{\sqrt{\dot{h}^2+h^2\dot{\bm{n}}^2}}, \\
 && b_2 = \frac{h|\dot{h}|}{\sqrt{\dot{h}^2+h^2\dot{\bm{n}}^2}}.
\ee
We find $\mathcal{L}O_2-b_2O_1=0$, which shows $d=3$ and $d_A=1$.
For $\dot{h}\ne 0$, the Krylov dimension $d=3$ equals the number of basis operators.
It is reduced to $d=2$ and $d_A=1$ for $\dot{h}=0$ where $b_2=0$.
We note that the eigenvalues of the Hamiltonian, $\pm h/2$,
are time independent when $\dot{h}=0$.

In any case, the dimension of the AGP is given by $d_A=1$.
We find the CD term
\be
 H_{\rm CD}=ib_0\alpha_1 O_1 =-i\frac{b_0b_1}{b_1^2+b_2^2}O_1
 = \frac{1}{2}\bm{n}\times \dot{\bm{n}}\cdot\bm{\Sigma}.
\ee
This result is consistent with the known 
result~\cite{Demirplak03, Demirplak05, Demirplak08, Berry09}.

\subsection{Driven harmonic quantum oscillator}

The driven harmonic oscillator is a workhorse
in nonequilibrium quantum dynamics and, not surprisingly, has played
a key role in the development of STAs~\cite{Chen10, Muga10}
and their experimental demonstration~\cite{Schaff10, An16}.
Although the dimension of Hilbert space is infinite, 
it is not difficult to treat the system analytically, since 
the system is a single-particle one, and the spectrum is discrete. 
In addition, its dynamics is described by a closed Lie algebra \cite{Mostafazadeh01}.

Consider the Hamiltonian
\be
 H(t)=\frac{1}{2m}P^2+\frac{1}{2}m\omega^2(t)[Q-q_0(t)]^2,
 \label{HO}
\ee
where $Q$ and $P$ are the position and momentum operators, respectively. 
Modulations of the time-dependent frequency $\omega$ induce expansions and
compressions, while transport processes are associated with variations
of the trap center $q_0$ \cite{Masuda10, Torrontegui11, An16}. 
We use the creation-annihilation operator representation
\be
 H(t) = \omega(t)\left(C^\dag(t) C(t)+\frac{1}{2}\right),
\ee
where
\be
 C(t)=\sqrt{\frac{m\omega(t)}{2}}[Q-q_0(t)]+i\sqrt{\frac{1}{2m\omega(t)}}P.
\ee
The CD term, in this case, is given by~\cite{Muga10, Ibanez12, Deffner14}
\be
 H_{\rm CD} &=& \dot{q}_0P
 -\frac{\dot{\omega}}{4\omega}\left[P(Q-q_0)+(Q-q_0)P\right] \no\\
 &=& i\dot{q}_0\sqrt{\frac{m\omega}{2}}\left(C^\dag-C\right)
 -i\frac{\dot{\omega}}{4\omega}\left(C^{\dag 2}-C^2\right). 
 \label{agpho}
\ee
Thus, the CD term involves a term proportional to the momentum operator,
the generator of spatial translations, and a second term proportional to
the squeezing operator, which is the generator of dilatations.

In the present case, we can explicitly calculate all the nested commutators.
As mentioned, using the basis operators $X$ is unnecessary.
We find 
\be
 \mathcal{L}^{2k}\dot{H} &=& 
 -\omega^{2k+1}\sqrt{\frac{m\omega}{2}}\dot{q}_0 (C^\dag+C)
 \\ && 
 +\omega^{2k+1}\frac{2^{2k-1}\dot{\omega}}{\omega}\left(C^{\dag 2}+C^2\right)
 +\delta_{k,0}\frac{\dot{\omega}}{\omega} H,\no
 \ee
for $k=0,1,\dots$, and 
\be
 \mathcal{L}^{2k-1}\dot{H}
 &=& -\omega^{2k}\sqrt{\frac{m\omega}{2}}\dot{q}_0 (C^\dag-C)
 \\ &&
 +\omega^{2k}\frac{2^{2(k-1)}\dot{\omega}}{\omega}\left(C^{\dag 2}-C^2\right),\no
\ee
for $k=1,2,\dots$.
These nested commutators involve a finite number of operators,
which determines the Krylov dimension.
It is given by
\be
 \begin{array}{ll}
 \mbox{$d=5$ and $d_A=2$} &
 \mbox{for $\dot{q}_0\ne 0$ and $\dot{\omega}\ne 0$,} \\
 \mbox{$d=3$ and $d_A=1$} &
 \mbox{for $\dot{q}_0= 0$ and $\dot{\omega}\ne 0$,} \\
 \mbox{$d=2$ and $d_A=1$} &
 \mbox{for $\dot{q}_0\ne 0$ and $\dot{\omega}=0$.}
 \end{array}
\ee
For $\dot{\omega}=0$, the eigenvalues of the Hamiltonian are
time independent, and the Krylov dimension is given by an even number.
The explicit form of the Krylov basis 
is given in Appendix~\ref{krylovforho}.

It is instructive to see how the exact AGP in Eq.~(\ref{agpho})
is obtained in the expansion.
In the case at $d_A=2$, the CD term is expanded as
$H_{\rm CD}=H_{\rm CD}^{(1)}+H_{\rm CD}^{(2)}$, and the first term 
$H_{\rm CD}^{(1)}=ib_0\alpha_1 O_1$ is given by 
\be
 H_{\rm CD}^{(1)}=r\dot{q}_0P-4r\frac{\dot{\omega}}{4\omega}
 \left[P(Q-q_0)+(Q-q_0)P\right],
\ee
where 
\be
 r=\frac{\dot{q}_0^2z_1^2+\frac{\dot{\omega}^2}{4\omega^2}z_2^2}
 {\dot{q}_0^2z_1^2+\frac{\dot{\omega}^2}{\omega^2}z_2^2},
\ee
with $z_1^2={\rm Tr}[\rho(H) P^2]$ and 
$z_2^2={\rm Tr}[\rho(H) \left(P(Q-q_0)+(Q-q_0)P\right)^2]$.
This result shows that each term of the expansion is dependent
on the definition of the inner product in Eq.~(\ref{innerprod}).

\subsection{STIRAP}

\begin{figure}[t]
\centering\includegraphics[width=1.\columnwidth]{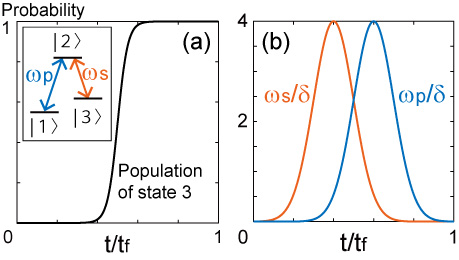}
\caption{
(a) Schematic view of STIRAP.
Three-level states are driven by two kinds of pulses, shown  in the inset, 
and the initial state $|1\rangle$ is adiabatically transferred to $|3\rangle$.
(b) A typical protocol for STIRAP.
We set $\omega_s(t) = \omega_0 \exp\left[-(t-t_1)^2/2\sigma^2\right]$ and 
$\omega_p(t) = \omega_0 \exp\left[-(t-t_2)^2/2\sigma^2\right]$ with
$\omega_0/\delta=4.0$, $\delta t_{\rm f}=100$, $t_1/t_{\rm f}=0.4$, 
$t_2/t_{\rm f}=0.6$, and $\sigma/t_{\rm f}=0.1$.
}
\label{fig03}
\end{figure}

As a practical application of three-level systems, we next discuss 
the stimulated Raman adiabatic passage (STIRAP)~\cite{Gaubatz90, Vitanov01}.
It is a method of population transfer between two states.
We introduce an additional state and apply two external pump fields
to the system.
The states are given by $|1\rangle$, $|2\rangle$, and $|3\rangle$, 
and we consider population transfer between $|1\rangle$ and $|3\rangle$.
The simplest STIRAP Hamiltonian is given by 
\be
 H(t)=\frac{1}{2}\left(\begin{array}{ccc} 0 & \omega_p(t) & 0 \\
 \omega_p(t) & 2\delta & \omega_s(t) \\ 0 & \omega_s(t) & 0 \end{array}\right).
\ee
A typical protocol is given in Fig.~\ref{fig03}.

Since we treat three-level systems, the number of independent operators 
is given by eight, except the identity operator.
We also see that the Hamiltonian is real symmetric and the $L$ matrix is
written as Eq.~(\ref{LM}).
Possible basis operators for $\rho(H)=1/2$ are given by 
\be
 X &=& \left\{\left(\begin{array}{ccc}
 0 & 1 & 0 \\ 1 & 0 & 0 \\ 0 & 0 & 0 
 \end{array}\right), \left(\begin{array}{ccc}
 0 & 0 & 0 \\ 0 & 0 & 1 \\ 0 & 1 & 0 
 \end{array}\right), 
 \left(\begin{array}{ccc}
 0 & 0 & 1 \\ 0 & 0 & 0 \\ 1 & 0 & 0 
 \end{array}\right),
 \right. \no\\ && \left.
 \left(\begin{array}{ccc}
 1 & 0 & 0 \\ 0 & -1 & 0 \\ 0 & 0 & 0 
 \end{array}\right), \frac{1}{\sqrt{3}}\left(\begin{array}{ccc}
 1 & 0 & 0 \\ 0 & 1 & 0 \\ 0 & 0 & -2 
 \end{array}\right)\right\}
\ee
and 
\be
 Y = \left\{\left(\begin{array}{ccc}
 0 & -i & 0 \\ i & 0 & 0 \\ 0 & 0 & 0 
 \end{array}\right), 
 \left(\begin{array}{ccc}
 0 & 0 & 0 \\ 0 & 0 & -i \\ 0 & i & 0 
 \end{array}\right), 
 \left(\begin{array}{ccc}
 0 & 0 & -i \\ 0 & 0 & 0 \\ i & 0 & 0 
 \end{array}\right)\right\}.
 \no\\ 
\ee
Using this basis, we find Eq.~(\ref{LM}) with 
\be
 M=i\left(\begin{array}{ccc}
 \delta & 0 & \omega_s/2 \\
 0 & -\delta & -\omega_p/2 \\
 \omega_s/2 & -\omega_p/2 & 0 \\
 \omega_p & -\omega_s/2 & 0 \\
 0 & \sqrt{6}\omega_s/2 & 0 
 \end{array}\right).
\ee
We note that the number of operators in $X$ can be reduced to four, 
as we can understand from the general discussions in the previous section.
Since they are not much different, we use the $5\times 3$ matrix here.

\begin{figure}[t]
\centering\includegraphics[width=1.\columnwidth]{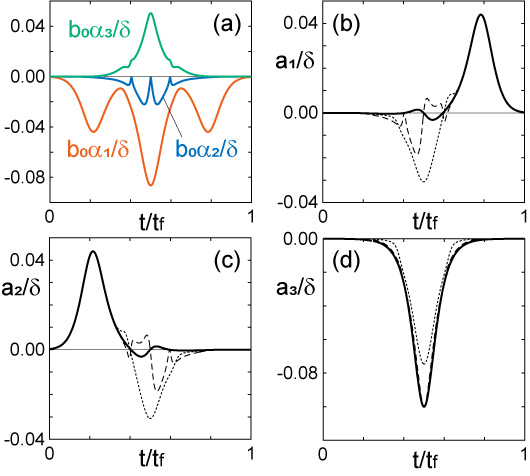}
\caption{
(a) Coefficients of the CD term $\alpha_k$ for STIRAP.
(b)--(d) $a_\mu^{(1)}$ (dotted curves), 
$a_\mu^{(1)}+a_\mu^{(2)}$ (dashed curves), and 
$a_\mu=\sum_{k=1}^3a_\mu^{(k)}$ (solid curves).
}
\label{fig04}
\end{figure}

We apply the Lanczos algorithm for the given $M$ matrix 
with the protocol in Fig.~\ref{fig03}(b) 
to calculate $\alpha_k$ shown in Fig.~\ref{fig04}(a).
The Krylov dimension is given by $d=7$.

The expansion is compared with the exact result~\cite{Chen10-2}
\be
 H_{\rm CD}(t)=-\dot{\phi}(t)\sin\theta(t) Y_1
 +\dot{\phi}(t)\cos\theta(t) Y_2-\dot{\theta}(t)Y_3, 
 \no\\
\ee
where $\theta(t) = \arctan(\omega_p(t)/\omega_s(t))$ and 
$\phi(t)=[\arctan(\sqrt{\omega_p^2(t)+\omega_s^2(t)}/\delta)]/2$.
In the Krylov method, the CD term is given by the form 
$H_{\rm CD}(t)=ib_0\sum_{k=1}^3\alpha_k O_{2k-1}$.
When we rewrite it as $H_{\rm CD}(t)=\sum_{\mu=1}^3 a_\mu Y_\mu$,
the coefficients are written as $a_\mu=\sum_{k=1}^3 a_\mu^{(k)}$, where 
\be
 a_\mu^{(k)}=ib_0\alpha_k (Y_\mu,O_{2k-1}),
\ee
and are plotted in Figs.~\ref{fig04}(b)--\ref{fig04}(d).
For short and large times, the adiabatic condition is approximately satisfied, 
and the CD term is well approximated by the first term of the Krylov expansion.

\section{Applications to many-body systems}
\label{sec:many}

The exact AGP or CD term is known for a limited number of many-body systems, and 
we expect that the Krylov method gives advantageous results that cannot be obtained 
from other methods.
For many-body systems, the required number of operators is large, 
and it is still a formidable task to find the exact CD term 
even in the present method.
In this section, we treat one-dimensional spin systems where 
the exact CD term is known for some examples.

\subsection{One-dimensional transverse Ising model}

\subsubsection{Exact result without a longitudinal magnetic field}
\label{SecTFQIM}
We first treat the Ising spin chain in a transverse field: 
\be
 H(t)=-\frac{v}{2}\left(\sum_{n=1}^{n_s} X_nX_{n+1}
 +g(t)\sum_{n=1}^{n_s} Z_n\right). \label{Hqising}
\ee
Many spins are aligned in a chain, and 
the number of spins is denoted by $n_s$.
We consider the periodic boundary condition, and 
the subscript is interpreted as ${\rm mod}\,n_s$.
We are interested in the large-$n_s$ limit where 
the system at $g=1$ shows a quantum phase transition~\cite{Sachdev11, Suzuki12}.

It is also known that the system is equivalent 
to the free fermion system~\cite{Lieb61,Katsura62}.
Then, the Hamiltonian is represented as an ensemble of two-level systems, 
and the CD term for each two-level system can be found from 
the result in the previous section.
Here, to study properties for many-body systems, we do not use the mapping
and treat the spin operators.

Under the setting $\rho(H)=1/(2^{n_s}n_s)$, we define orthonormalized operators 
\be
 && M=\sum_{n=1}^{n_s} Z_n, \\
 && V_k^X=\sum_{n=1}^{n_s}X_nZ_{n+1}\cdots Z_{n+k-1}X_{n+k}, \\
 && V_k^Y=\sum_{n=1}^{n_s}Y_nZ_{n+1}\cdots Z_{n+k-1}Y_{n+k}, \\
 && W_k=\frac{1}{\sqrt{2}}\sum_{n=1}^{n_s}\left(X_nZ_{n+1}\cdots Z_{n+k-1}Y_{n+k}
 \right. \no\\ && \left.
 +Y_nZ_{n+1}\cdots Z_{n+k-1}X_{n+k}\right),
\ee
with $k=1,2,\dots, n_s-1$.
These operators take bilinear forms in fermion operators and 
can be used as basis operators.
Since the Hamiltonian commutes with $(-1)^P=\prod_{n=1}^{n_s}Z_n$, 
the number of independent operators is reduced.
We have the relations 
$V_{n_s-k}^X=-(-1)^P V_k^Y$, $V_{n_s-k}^Y=-(-1)^P V_k^X$, and $W_{n_s-k}=(-1)^P W_k$.
Then, the subscript takes $k=1,2,\dots,n_s/2$ for even $n_s$.

\begin{figure}[t]
\centering\includegraphics[width=1.\columnwidth]{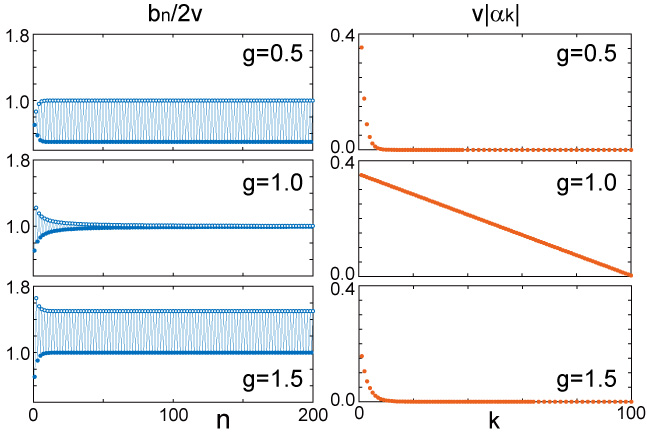}
\caption{
The Lanczos coefficients (left) and 
the coefficients of the CD term (right)
for the one-dimensional transverse Ising model in Eq.~(\ref{Hqising}) 
at $n_s=200$.
For the Lanczos coefficients, we denote 
$b_1, b_3, \dots$ by the filled circle 
and  $b_2, b_4, \dots$ by the open circle.
}
\label{fig05}
\end{figure}

Acting $\mathcal{L}$ to these operators, we obtain 
\be
 && \mathcal{L}M=\sqrt{2}iv W_1, \\
 && \mathcal{L}V_k^X=\sqrt{2}iv\left(W_{k-1}-gW_k\right), \\
 && \mathcal{L}V_k^Y=\sqrt{2}iv\left(-W_{k+1}+gW_k\right), \\
 && \mathcal{L}W_k=\sqrt{2}iv\left[
 V_{k-1}^Y-V_{k+1}^X+g(V_{k}^X-V_{k}^Y)-\delta_{k,1}M\right]. \no\\
\ee
Since the Hamiltonian is real symmetric, the nested commutators at odd order
involve only $W_k$.
In Appendix \ref{krylovforqising}, we show 
\be
 O_{2k-1}=(-1)^{k}iW_k
\ee
and 
\be
 && b_1^2+b_2^2=b_3^2+b_4^2=\cdots=b_{d-2}^2+b_{d-1}^2
 =4v^2(1+g^2), \no\\ \\
 && b_2b_3=b_4b_5=\cdots=b_{d-3}b_{d-2}=4v^2g.
\ee
The Krylov dimension is odd in this case.
Using the result $b_0=\sqrt{n_s}v\dot{g}(t)/2$ and $b_1=\sqrt{2} v$, 
where we assume $v>0$ and $\dot{g}(t)>0$, we can calculate 
all of the Lanczos coefficients.
We plot the Lanczos coefficients in the left in Fig.~(\ref{fig05})
for several values of $g$.
The asymptotic forms satisfy $b_{2k}\sim 2v\ge b_{2k-1}\sim 2vg$ for $g\le 1$ 
and $b_{2k}\sim 2vg\ge b_{2k-1}\sim 2v$ for $g\ge 1$.
This result implies that we can generally find a phase transition point 
from the Lanczos coefficients.

The coefficients of the CD term are obtained from Eq.~(\ref{alphaoddd}). 
Since the matrix in the equation has the diagonal components 
$4v^2(1+g^2)$ and the nonzero off-diagonal components $4v^2g$, 
we can invert the matrix by using the discrete Fourier transformation.
The details are given in Appendix \ref{krylovforqising}.
We obtain 
\be
 && (-1)^{k-1}b_0\alpha_k = \dot{g}\sqrt{\frac{n_s}{2}}\frac{1}{2(d_A+1)}
 \no\\ && \times
 \sum_{l=1}^{d_A}
 \frac{\sin\frac{\pi l}{d_A+1}}{1+g^2-2g\cos\frac{\pi l}{d_A+1}}
 \sin\frac{k\pi l}{d_A+1}.
  \label{aqising}
\ee
The CD term is given by $H_{\rm CD}=\sum_{k=1}^{d_A} (-1)^{k-1}b_0\alpha_k W_k$
with $d_A=n_s/2$.
This result entirely agrees with that in Refs.~\cite{delCampo12, Takahashi13,Damski14}.
Since the Krylov basis at odd order in the present case is given by a simple 
form $W_k$, it is natural to find the same result without using the Krylov method.

We plot $\alpha_k$ for several values of $g$ 
in the right in Fig.~\ref{fig05}.
For $g\ne 1$, the Lanczos coefficients at even order are larger 
than those at odd order; we see from Eq.~(\ref{alphaoddd2}) 
that $\alpha_k$ is a decreasing function in $k$.
Then, we find that $\alpha_k$ rapidly decreases, which means that 
few-body interactions become the dominant contributions to the CD term.
It was discussed in Refs.~\cite{delCampo12, Damski14} that 
$\alpha_k\sim g^{k-1}$ for $|g|< 1$
and $\alpha_k\sim g^{-k-1}$ for $|g|> 1$
at the thermodynamic limit $n_s\to\infty$.
For $g=1$, the Lanczos coefficients take a constant value, and 
$\alpha_k$ decreases slowly as a function of $k$.
At $n_s\to\infty$, an infinite number of operators 
contributes to the CD term as discussed in Ref.~\cite{delCampo12}.

\subsubsection{Approximate result with a longitudinal magnetic field}

The free fermion representation is possible only when the direction
of the magnetic field is perpendicular to the direction of
the interaction operator.
Here, we treat the nonintegrable Hamiltonian 
\be
 H(t) &=& g(t)\left(-v\sum_{n=1}^{n_s}Z_nZ_{n+1}-h\sum_{n=1}^{n_s}Z_n\right)
 \no\\
 && +(1-g(t))\left(-\gamma\sum_{n=1}^{n_s}X_n\right),
 \label{qa}
\ee
with $g(t)=t/t_{\rm f}$.
This is the standard form of the quantum annealing Hamiltonian and 
a test bed for universal dynamics of quantum phase transitions with a bias \cite{Rams19}.
We consider the time evolution from 
the ground state of $H(0)$ toward that of $H(t_{\rm f})$.
The nonadiabatic effect makes the system deviate from the instantaneous
ground state, and we apply the CD term to prevent the transition.

It is a complex problem to find and implement
the exact CD term for this Hamiltonian, 
and we consider a restricted number of operators.
As we discussed in the above example, each term of $H_{\rm CD}(t)$ involves
an odd number of Pauli $Y$ operators.
We keep the $Y$ basis up to two-body operators 
\be
 Y &=& \left(\sum_{n=1}^{n_s}Y_n,\ 
 \frac{1}{\sqrt{2}}\sum_{n=1}^{n_s}\left(Y_nZ_{n+1}+Z_nY_{n+1}\right),\ 
 \right.\no\\
 && \frac{1}{\sqrt{2}}\left.
 \sum_{n=1}^{n_s}\left(Y_nX_{n+1}+X_nY_{n+1}\right)\right),
\ee
with $\rho(H)=1/(2^{n_s}n_s)$.
We act with ${\cal L}$ on these operators to construct the $M$ matrix.
Using the $X$ basis 
\be
 && X = \left(\sum_{n=1}^{n_s}Z_n,\ \sum_{n=1}^{n_s}X_n,\ 
 \sum_{n=1}^{n_s}Z_nZ_{n+1},\ 
 \right. \no\\
 && \sum_{n=1}^{n_s}X_nX_{n+1},\ \sum_{n=1}^{n_s}Y_nY_{n+1}, \no\\
 && \frac{1}{\sqrt{2}}\sum_{n=1}^{n_s}\left(Z_nX_{n+1}+X_nZ_{n+1}\right),\
 \sum_{n=1}^{n_s}Z_{n-1}X_nZ_{n+1}, \no\\
 && \frac{1}{\sqrt{2}}\sum_{n=1}^{n_s}
 \left(Z_{n-1}X_nX_{n+1}+X_{n-1}X_{n}Z_{n+1}\right), \no\\
 && \frac{1}{\sqrt{2}}\left.
 \sum_{n=1}^{n_s}\left(Z_{n-1}Y_nY_{n+1}+Y_{n-1}Y_{n}Z_{n+1}\right)\right), 
\ee
we can construct the $M$ matrix with the size $9\times 3$
and apply the Lanczos algorithm to calculate the approximate CD term.
We note that the size of the $M$ matrix can be reduced to $3\times 3$, 
keeping the result unchanged.
The Krylov dimension is given by $d=7$.

\begin{figure}[t]
\centering\includegraphics[width=1.\columnwidth]{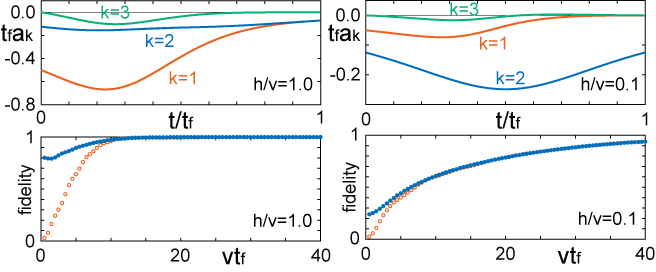}
\caption{
The time evolution for the quantum annealing Hamiltonian in Eq.~(\ref{qa}).
We set $N=6$, $\gamma/v=1.0$, and $h/v=1.0$ for the left and
$h/v=0.1$ for the right. 
In the upper, we plot $(a_1(t),a_2(t),a_3(t))$
for the approximate CD term
$H_{\rm CD}(t)=a_1(t)Y_1+\sqrt{2}a_2(t)Y_2+\sqrt{2}a_3(t)Y_3$.
In the lower, we plot the fidelity $f$ for the time evolutions
with $H(t)$ (red open circle) and
with $H(t)+H_{\rm CD}(t)$ (blue filled circle).
}
\label{fig06}
\end{figure}

We calculate the approximate CD term and
numerically solve the time evolution with and without the approximate CD term
by setting the initial state as the ground state of $H(0)$
for the system size $N=6$.
In Fig. ~\ref{fig06}, we plot the approximate CD term as a function of $t$
and the fidelity
\be
 f=|\langle \psi_{\rm gs}(t_{\rm f})|\psi(t_{\rm f})\rangle|^2,
\ee
as a function of the annealing time $t_{\rm f}$.
Here, $|\psi(t)\rangle$ represents the time-evolved state
with or without the approximate CD term, and 
$|\psi_{\rm gs}(t)\rangle$ represents the instantaneous ground state of $H(t)$.
$f$ is close to unity when the adiabaticity condition is satisfied.

We numerically confirm that the present method gives the same result
as the variational method, which is expected from the general discussion.
When $h/v$ takes a large value, the ground-state energy is well separated from
the other ones, and we find a large fidelity.
As $h/v$ decreases to zero, the fidelity becomes smaller
and finding the exact ground state becomes difficult 
even with the approximate CD term in use.
As we see from the figure, many-body terms in the CD term become important
for small $h$, and we require higher-order terms to improve the result
by the approximate CD driving.

\subsection{One-dimensional XX model}

\subsubsection{Krylov method}

In the example of the Hamiltonian in Eq.~(\ref{Hqising}), 
$O_{2k-1}$ incorporates $k+1$-body interactions only,
which is a specific property for the transverse Ising model.
Here, to study more complicated situations, 
we treat the one-dimensional isotropic XY model (XX model) 
with the open boundary condition 
\be
 H(t) &=& \frac{1}{2}\sum_{n=1}^{n_s-1} v_n(t)(X_nX_{n+1}+Y_nY_{n+1})
 \no\\
 && +\frac{1}{2}\sum_{n=1}^{n_s} h_n(t)Z_n.
 \label{Hxx}
\ee
This model is also known to be equivalent to a free fermion model.
However, the coefficients are dependent on the site index $n$ and 
it is generally a difficult task to find the exact CD term.

The mapping to a bilinear form in fermion operators denotes 
that we can find a closed algebra within a limited number of 
operators.
For $\rho(H)=1/2^{n_s}$, we define 
\be
 && V_n^k=\frac{1}{\sqrt{2}}
 \left(X_nZ_{n+1}\cdots Z_{n+k-1}X_{n+k}
 \right. \no\\ && \left.
 +Y_nZ_{n+1}\cdots Z_{n+k-1}Y_{n+k}\right), \\
 && W_n^k=\frac{1}{\sqrt{2}}\left(X_nZ_{n+1}\cdots Z_{n+k-1}Y_{n+k}
 \right. \no\\ && \left.
 -Y_nZ_{n+1}\cdots Z_{n+k-1}X_{n+k}\right),
 \label{VWeqs}
\ee
with $n=1,2,\dots,n_s$ and $k=1,2,\dots,n_s-n$.
Since the Hamiltonian is real symmetric, we define 
$X=(\{Z_n\},\{V_n^k\})$ and $Y=(\{W_n^k\})$ to construct 
the $M$ matrix with the size $n_s(n_s+1)/2\times n_s(n_s-1)/2$.
We have
\be
 && \mathcal{L}W_n^k=-i(h_{n+k}-h_n)V_n^k
 -iv_{n-1}V_{n-1}^{k+1}-iv_{n}V_{n+1}^{k-1} 
 \no\\ &&
 +iv_{n+k-1}V_{n}^{k-1}+iv_{n+k}V_{n}^{k+1}
 \no\\ &&
 +\delta_{k,1}\sqrt{2}iv_n (Z_{n+1}-Z_n).
\ee

\begin{figure}[t]
\centering\includegraphics[width=1.\columnwidth]{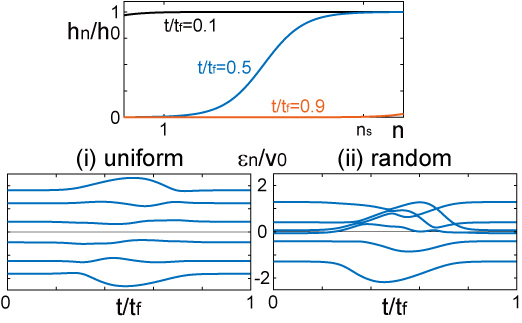}
\caption{
The protocol for the XX model in Eq.~(\ref{Hxx}).
The upper represents $h_n(t)$.
We take $h_n(t)=h_0(1+\tanh f_n(t))/2$ and 
$f_n(t)=n-1+x_0-(n_s-1+2x_0)t/t_{\rm f}$ with $x_0=4$ and $n_s=6$.
The lower represents the instantaneous eigenvalues of the Hamiltonian
for the uniform case (i) and the random case (ii) with $h_0/v_0=2.0$.
In the following calculations, we take $v_0t_{\rm f}=100$.
}
\label{fig07}
\end{figure}

For the Hamiltonian coefficients 
$\{v_n(t)\}_{n=1}^{n_s-1}$ and $\{h_n(t)\}_{n=1}^{n_s}$,
we consider an annealing protocol.
We take $v_n$ as a constant value and 
consider the two cases: (i) uniform distribution $v_n=v_0>0$ and 
(ii) random distribution $v_n=v_0r_n$, where $r_n$ is 
a uniform random number with $r_n\in [-1,1]$.
For a given set of $\{v_n\}$, 
we change the magnetic field $h_n(t)$ from $h_n(0)=h_0 \gg |v_n|$ 
to $h_n(t_{\rm f})=0$.
We take $h_n(t)=h_0(1+\tanh f_n(t))/2$ 
with $f_n(t)=n-1+x_0-(n_s-1+2x_0)t/t_{\rm f}$.
The system becomes adiabatic for large $t_{\rm f}$.
The parameter $x_0$ takes a large value so that the conditions at 
$t=0$ and $t=t_{\rm f}$ are satisfied.
We plot the protocols used in the following calculations in 
the upper in Fig.~\ref{fig07}.
The instantaneous eigenvalues for (i) and (ii) are, respectively, 
plotted in the lower in Fig.~\ref{fig07}.
The Hamiltonian commutes with $M=\sum_{n=1}^{n_s}Z_n$, and 
we consider the block with $M=n_s-2$.
The Hamiltonian takes a tridiagonal form, and 
the size of the matrix is given by $n_s$.

\begin{figure}[t]
\centering\includegraphics[width=1.\columnwidth]{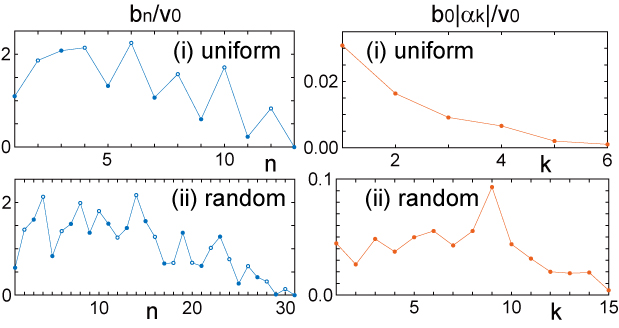}
\caption{
The Lanczos coefficients $b_n$ and the coefficients of the CD term, 
$\alpha_k$, for the XX model with $n_s=6$ at $t/t_{\rm f}=0.5$.
The top are results for case (i), uniform distribution, 
and the bottom are for (ii), random distribution.
For the Lanczos coefficients, we denote 
$b_1, b_3, \dots$ by the filled circle  
and  $b_2, b_4, \dots$ by the open circle.
}
\label{fig08}
\end{figure}

Although it is not difficult to implement the Lanczos algorithm numerically 
for a considerably large value of $n_s$, 
we here take $n_s=6$ to keep a good visibility of the plotted points.
In Fig.~\ref{fig08}, we display the Lanczos coefficients 
and the coefficients of the CD term for a fixed $t$.
The Lanczos coefficients show an oscillating behavior.
For the uniform case (i), 
the even order tends to be larger than the odd order, and 
correspondingly $|\alpha_k|$ shows a decreasing behavior.
For the random case (ii), we observe a complicated behavior denoting 
that the higher-order contributions of the Lanczos expansion are important.

\begin{figure}[t]
\centering\includegraphics[width=1.\columnwidth]{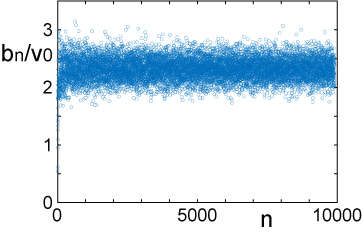}
\caption{The Lanczos coefficients of the XX model with $n_s=100$.
We consider the random case (ii) and take $t/t_{\rm f}=0.5$.
}
\label{fig09}
\end{figure}

The typical behavior of $b_n$ for a large $n_s$ is shown in Fig.~\ref{fig09}.
For any choice of parameters, we observe a flat band, which is considered 
to be a property for ``noninteracting'' systems.

\begin{figure}[t]
\centering\includegraphics[width=1.\columnwidth]{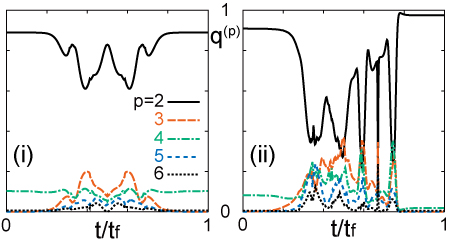}
\caption{Time dependence of the norm fraction
$\{q^{(p)}(t)\}_{p=2}^{n_s}$ in Eq. (\ref{normratios}), 
weighting the contribution of the $p$-body interactions to the CD term, 
for the XX model with $n_s=6$. 
Many-body CD terms are shown to be necessary 
when the energy levels change significantly along the driving protocol.
}
\label{fig10}
\end{figure}

To study the properties of the obtained CD term, we next decompose it.
The Lanczos basis at odd order is written as 
\be
|\theta_{2k-1}\rangle=\left(\begin{array}{cc}
|\theta_{2k-1}^{(2)}\rangle \\
|\theta_{2k-1}^{(3)}\rangle \\
\vdots \\
|\theta_{2k-1}^{(n_s)}\rangle
\end{array}\right),
\ee
where $|\theta_{2k-1}^{(p)}\rangle$ has $n_s+1-p$ components and
the corresponding Krylov basis involves $p$-body interactions.
We decompose the norm of the CD term as 
$\langle H_{\rm CD}|H_{\rm CD}\rangle
=\sum_{p=2}^{n_s}\langle H_{\rm CD}^{(p)}|H_{\rm CD}^{(p)}\rangle$ 
where $|H_{\rm CD}^{(p)}\rangle
=-b_0\sum_{k=1}^{d_A}\alpha_k|\theta_{2k-1}^{(p)}\rangle$, 
and define the norm fraction 
\be
 q^{(p)}=\frac{\langle H_{\rm CD}^{(p)}|H_{\rm CD}^{(p)}\rangle}
 {\langle H_{\rm CD}|H_{\rm CD}\rangle}=\frac{\sum_{k=1}^{d_A}\alpha_k^2
 \langle\theta_{2k-1}^{(p)}|\theta_{2k-1}^{(p)}\rangle}
 {\sum_{k=1}^{d_A}\alpha_k^2},
 \label{normratios}
\ee
which weights the contribution of the $p$-body term.
This is a function of $t$ and is plotted for each value of $p$ 
in Fig.~\ref{fig10}.
We can understand from the comparison between 
the energy levels in Fig.~\ref{fig07} and $q^{(p)}$ in Fig.~\ref{fig10} 
that the many-body interaction terms cannot be neglected 
when the energy levels significantly change as a function of $t$.

\begin{figure}[t]
\centering\includegraphics[width=1.\columnwidth]{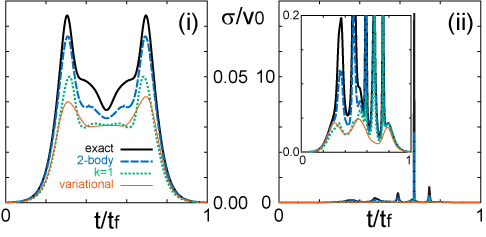}
\caption{
The norm of the CD term for the XX model at $n_s=6$.
Bold solid curves "exact" (black) represent 
$\sigma=\sqrt{\langle H_{\rm CD}|H_{\rm CD}\rangle}$.
We use the replacement $H_{\rm CD}\to H_{\rm CD}^{(2)}$ 
for dashed curves "2-body" (blue),
$H_{\rm CD}\to ib_0\alpha_1O_1$ for dotted curves "$k=1$" (green), and  
$H_{\rm CD}\to i\alpha_1^{\rm nc}\mathcal{L}\dot{H}$ 
for thin solid curves "variational" (red) where $\alpha_1^{\rm nc}$ 
is obtained from the variational method.
}
\label{fig11}
\end{figure}

It is generally understood from Eq.~(\ref{Asp}) that 
the CD term gives a large contribution 
when some of the energy levels are close to each other.
We calculate the amplitude of the CD term 
$\langle H_{\rm CD}|H_{\rm CD}\rangle$, 
which is plotted in Fig.~\ref{fig11}.
As we have discussed, the CD term is decomposed as 
$H_{\rm CD}=\sum_p H_{\rm CD}^{(p)}$ and 
$H_{\rm CD}=ib_0\sum_k\alpha_k O_{2k-1}$.
In the same figure, we plot the result where the first term of 
the expansion is kept for each decomposition.
We also plot the first contribution of the approximate CD 
term $H_{\rm CD}^{\rm nc}$ obtained from the expansion in Eq.~(\ref{var}).
The coefficient $\alpha_1^{\rm nc}$ is obtained 
from the minimization condition of 
${\rm Tr}[(\dot{H}-i\mathcal{L}H_{\rm CD}^{\rm nc})^2]$~\cite{Sels17, Claeys19}.
The result implies that the approximate CD term underestimates 
an abrupt growth of the CD term.

We note that this result does not necessarily lead to 
the failure of the approximation method.
The CD term is independent of the choice of the initial condition
of the time evolution.
In the present examples, as we see from Fig.~\ref{fig07},  
the energy level crossings occur at higher energy levels.
The ground-state level is isolated from the other levels, and 
we can expect that a large amplitude of the CD term is not required for 
the control of the ground state. 
The situation is, in this sense, opposite to that across 
a quantum phase transition, discussed in Sec. \ref{SecTFQIM}, 
in which the gap between the ground state and the first excited state closes, 
and the norm of the CD exhibits a singularity \cite{delCampo12,Damski14}.

In the present analysis, we set the measure $\rho(H)$ 
in the inner product as $\rho(H)=1/2^{n_s}$. 
When we control a state with an energy level well separated from the other levels,
it is reasonable to consider a weighted measure 
such as the Gibbs--Boltzmann distribution $\rho(H)\propto e^{-\beta H}$.
Although the exact CD term is independent of the measure, 
each term of the expansion $H_{\rm CD}=ib_0\sum_{k=1}^{d_A}\alpha_k O_{2k-1}$ 
is dependent on it.
Therefore, when we consider a truncation approximation in the Krylov expansion,
the choice of the measure strongly influences the result.
Since a nontrivial choice of the measure makes 
the calculation of the inner product difficult, 
it is practically an interesting problem 
to find a convenient form of the measure.
For the ground state, it is tempting to take the limit 
$\beta\to \infty$ for $\rho(H)\propto e^{-\beta H}$.
However, the measure is not positive definite in this limit, and 
we find unexpected behavior, such as the vanishing of 
the Lanczos coefficient $b_n$ for $n<d-1$.

\subsubsection{Toda equation}

It is known that the exact CD term is analytically obtained when 
the coefficients of the Hamiltonian satisfy the Toda equations \cite{Okuyama16} 
\be
 && \dot{h}_n(t)=2[v_n^2(t)-v_{n-1}^2(t)], \\
 && \dot{v}_n(t)=v_n(t)[h_{n+1}(t)-h_n(t)].
\ee
The CD term is then given by 
\be
 H_{\rm CD}(t)=\frac{1}{\sqrt{2}}\sum_{n=1}^{n_s-1}v_n(t)W_n^1,
\ee
with $W_n^1$ as in Eq. (\ref{VWeqs}).
This term satisfies $\dot{H}-i\mathcal{L}H_{\rm CD}=0$, which means that 
the instantaneous eigenvalues of the Hamiltonian are time independent. 

Despite this simplicity, 
the corresponding Krylov expansion is generally involved and is required at each time.
The time evolution of the Hamiltonian is reflected only in 
the choice of the initial basis $b_0O_0=\dot{H}$, and 
the expansion is essentially insensitive to the choice.
Except for the special cases discussed below, 
we find that the Krylov dimension is as large as 
the number of odd basis elements $n_s(n_s-1)/2$.
Highly nontrivial cancellations should be observed 
when we calculate $\alpha_k$ to give the result with $q^{(p)}=\delta_{p,2}$.

As a very special case, we can find the result with $d=2$ and $d_A=1$ 
when the coefficients are written as 
\be
 && h_n(t)=-\frac{2h_1}{n_s-1}\left(n-\frac{n_s+1}{2}\right)\sin\theta(t), \\
 && v_n^2(t)=\frac{n(n_s-n)}{(n_s-1)^2}h_1^2\cos^2\theta(t).
\ee
Using the Toda equations, 
we obtain that a single equation describes the time evolution 
\be
 \frac{\dot{\theta}(t)}{\cos\theta(t)}=\frac{2h_1}{n_s-1}.
\ee
For a given $h_1$ and a initial condition $\theta(0)$, 
the coefficients evolve, keeping the equidistant of $h_n(t)$ and
a quadratic form of $v_n(t)$.
In this case, we find that the first-order term in the Krylov expansion is 
proportional to the exact CD term, i.e.,  
\be
 b_0b_1O_1=
 \mathcal{L}\dot{H}(t)=\frac{i}{\sqrt{2}}\left(\frac{2h_1}{n_s-1}\right)^2
 \sum_{n=1}^{n_s-1}v_n(t)W_n^1.
\ee
Since $\mathcal{L}O_1-b_1O_0=0$, the expansion terminates at this order, 
and we obtain a simple result with $d=2$.

It was discussed that the present choice of parameters saturates the operator 
speed limit, i.e., the quantum speed limit in unitary operator flows~\cite{Hornedal23}.
The nested commutators span the Krylov space within a limited number of operators.

\section{Discussion and Summary}
\label{sec:summary}

The use of the integral representation of the CD term introduced 
in Ref.~\cite{Claeys19} has eased the study of STA in many-body systems by 
removing the requirement for the exact diagonalization of 
the instantaneous system Hamiltonian. 
In its place, the CD term can be expressed as a series of nested commutators that 
follows from the Baker--Campbell--Hausdorff formula. 
The coefficients in such an expansion can be determined through 
a variational principle~\cite{Sels17}.

In this work, we have introduced Krylov subspace methods 
to provide an exact expression of the CD term. 
The Krylov algorithm identifies an operator basis for the terms 
generated in the series by nested commutators, 
along with the set of Lanczos coefficients. 
Using these two ingredients, we have provided an exact closed-form expression of 
the CD term, circumventing the need for a variational approach. 
When the dimension of the Hilbert space is finite, 
the series by the Krylov basis is finite, 
which is in contrast to the expansion using nested commutators.

We have shown the applicability of our method in the paradigmatic models
in which the CD term admits an exact closed-form solution. 
This includes single-particle systems such as two- and three-level systems and 
the driven quantum oscillator.
Although the applications of the present method can be laborious, 
the implementation of our approach in these systems is a straightforward task, 
applicable to any Hamiltonian with no special symmetry.
We have further applied our formalism to a variety of quantum spin chain models 
encompassing the cases in which the system is integrable, nonintegrable, and disordered. 
Specifically, we applied the construction of the CD term in Krylov space to 
the one-dimensional transverse-field quantum Ising model as 
a paradigmatic instance of an integrable and solvable quasi-free fermion Hamiltonian. 
We have further demonstrated our approach in the presence of 
a longitudinal symmetry-breaking bias field that breaks integrability. 
A similar study is possible for the XX model 
where the free fermion representation is available.
However, in that case, the site-dependent couplings make 
the explicit construction of the CD term by the standard method difficult,
except for the special case when the coupling constants are varied in time 
according to a Toda flow, and the resulting CD term takes a simple local form.
We have explicitly constructed the CD term for general and disordered couplings, 
and the result was compared to 
the approximation method~\cite{Sels17, Claeys19}.

The main task of the Krylov algorithm is to construct 
the basis operators for the minimal subspace in which the dynamics unfolds and 
to determine the associated Lanczos coefficients by iterations.
The CD term is constructed as a series involving only the odd-order operators of 
the Krylov basis,
with the corresponding coefficients in this compact expansion being determined 
in terms of the Lanczos coefficients.
These properties imply that we can find some implications by comparing 
the Lanczos coefficients of odd order and those of even order.
As we see from Figs.~\ref{fig05} and \ref{fig08}, 
the Lanczos coefficients typically show an oscillating behavior.
When the coefficients at even order are larger than those at odd order,
the corresponding coefficients of the CD term show a decaying behavior, 
and the CD term is well approximated by the first several terms of the expansion.
We also find that the Krylov dimension is even when 
the instantaneous eigenstates of the original Hamiltonian are time independent.
Thus, we can directly find the dynamical properties of the system 
from the Krylov algorithm.

The Krylov expansion is dependent on the choice of the inner product.
Although the CD term is independent of the choice,
each term in Eq.~(\ref{agpk}) is sensitive to it, a feature that can be relevant 
when considering truncating approximations.
It is an interesting problem to find a proper choice 
depending on the situation to treat.
We can also consider the truncation of the Krylov subspace.
To this end, it suffices to restrict the basis operators  
and to construct the $L$ matrix in the truncated space.
Then, we can apply the Lanczos algorithm to find an approximate CD term. 
We note that even in that case 
the coefficients of the CD term are obtained without using any variational procedures,
which gives a different result from the variational method.

Our primary emphasis has been on exact and analytical results 
formulating the CD term in Krylov space. 
In addition, there exist powerful numerical algorithms that largely simplify 
the computation of the Lanczos coefficients, used in our methodology. 
These are well established in the literature on Krylov subspace methods and 
numerical analysis~\cite{Liesen12}. 
They are further available in popular software and numerical routines. 
They hold for any matrix of Hessenberg form and replace 
the Gram-Schmidt diagonalization by the use of Householder reflections, 
making the implementation numerically stable and computationally efficient.

Our work offers an interesting prospect to improve state-of-the-art quantum algorithms 
by combining the formulation of the CD term in Krylov space with 
the digital approach for quantum simulation. 
This approach may prove advantageous over the current formulation relying on variational 
methods~\cite{Hegade21,Hegade21factorization,Hegade21Portfolio,Chandarana22,
Hegade22DCQO,Chandarana22-2}, 
suggesting the need to generalize the error scaling in digitizing 
the CD term~\cite{Hatomura23} to Krylov space.

In summary, we have proposed a technique for constructing the CD term exploiting 
the Krylov operator space.
The method is flexibly applied to systems with many degrees of freedom and 
can be a powerful general method for understanding the dynamical properties
of the system in control.
Suppression of nonadiabatic transitions in discrete systems 
with many degrees of freedom is one of the dominant problems 
in quantum technologies, such as quantum simulation and quantum computing, 
and we hope that our method will be an efficient technique inspiring further studies. 

{\it Note added.} -- Recently, related results were reported 
in Ref.~\cite{bhattacharjee2023lanczos}. 

\section*{Acknowledgments}

We are grateful to Ruth Shir for useful discussions. 
We acknowledge financial support from
the project QuantERA II Programme STAQS project that has
received funding from the European Union’s Horizon 2020 research and 
innovation program under Grant Agreement No. 16434093. 
K.T. was supported by 
JSPS KAKENHI Grants No. JP20H01827 and No. JP20K03781.

\appendix
\section{Derivation of Eq.~(\ref{AA})}
\label{deriveAA}

In Sec.~\ref{owf}, we discussed the relation between the AGP
and the operator wave function $|\varphi(s)\rangle$.
The wave function satisfies
$\partial|\varphi(s)\rangle=B|\varphi(s)\rangle$
with the matrix $B$ in Eq.~(\ref{b}) and the initial condition 
$|\varphi(0)\rangle=(1,0,0,\dots)^{\rm T}$.
Since $B$ is independent of $s$, we can solve the differential equation
by the standard method for stationary states.
Since $iB$ is Hermitian, the eigenvalue equation 
\be
 iB|\omega_n\rangle= \omega_n|\omega_n\rangle
\ee
is solved to find a real eigenvalue $\omega_n$.
We also find that $B$ anticommutes with 
$Z={\rm diag}\,(1,-1,1,-1,\dots)$, which gives the relation
\be
 iBZ|\omega_n\rangle= -\omega_nZ|\omega_n\rangle.
\ee
This identity shows that, 
for the eigenstate $|\omega_n\rangle$ with the eigenvalue $\omega_n$,
$Z|\omega_n\rangle$ represents an eigenstate with
the eigenvalue $-\omega_n$.
We introduce the  decomposition $|\omega_n\rangle=|\omega_n^+\rangle+|\omega_n^-\rangle$, 
where 
\be
 |\omega_n^{\pm}\rangle=\frac{1\pm Z}{2}|\omega_n\rangle.
\ee
Then, the associated orthonormality relations 
$\langle\omega_m|\omega_n\rangle=\delta_{m,n}$ and 
$\langle -\omega_m|\omega_n\rangle=0$ for $\omega_{m,n}>0$ yield
\be
 \langle\omega_m^+|\omega_n^+\rangle=\langle\omega_m^-|\omega_n^-\rangle
 =\frac{1}{2}\delta_{m,n}.
\ee
We also discuss in the main body of the paper that 
the zero-eigenvalue state $|\phi\rangle$ exists only when the dimension is odd. 

Using the eigenstates discussed above, we can generally write 
\be
 |\varphi(s)\rangle &=& \sum_{n(\omega_n>0)}
 \left(e^{-i\omega_ns}
 +e^{i\omega_ns}Z\right)
 |\omega_n\rangle\langle\omega_n|\varphi(0)\rangle
 \no\\ &&
 +|\phi\rangle\langle\phi|\varphi(0)\rangle. 
\ee
The last term exists only for odd dimensions.
We use this representation for the integral form in Eq.~(\ref{phia}).
The integration over $s$ is performed to give 
\be
 (-1)^k\alpha_k = \sum_{n(\omega_n>0)}\frac{2}{i\omega_n}
 \langle 2k-1|\omega_n\rangle\langle\omega_n|0\rangle,
\ee
where $\langle n|\psi\rangle$ denotes 
$\psi_n$ for a vector $|\psi\rangle=(\psi_0,\psi_1,\dots)^{\rm T}$.
We also use $\langle 2k-1|Z=-\langle 2k-1|$ and $\langle 2k-1|\phi\rangle =0$
to obtain this result.
Taking the square and the sum over the index, we obtain 
\be
 \sum_k \alpha_k^2
 &=& \sum_k\sum_{m(\omega_m>0)}\sum_{n(\omega_n>0)}
 \frac{4}{\omega_m\omega_n}
 \no\\ && \times
 \langle 0|\omega_m\rangle\langle\omega_m|2k-1\rangle
 \langle 2k-1|\omega_n\rangle\langle\omega_n|0\rangle
 \no\\
 &=& \sum_{m(\omega_m>0)}\sum_{n(\omega_n>0)}
 \frac{4}{\omega_m\omega_n}
 \langle 0|\omega_m\rangle\langle\omega_m^-|\omega_n^-\rangle
 \langle\omega_n|0\rangle
 \no\\
 &=& \sum_{n(\omega_n\ne 0)}
 \frac{1}{\omega_n^2}
 \langle 0|\omega_n\rangle\langle\omega_n|0\rangle.
\ee
Thus, we find Eq.~(\ref{AA}).
Since $iB$ is related to the matrix $T$ in Eq.~(\ref{T})
by a unitary transformation, we can also write 
\be
 \sum_k \alpha_k^2 = \langle 0|(QTQ)^{-2}|0\rangle.
\ee

\section{Krylov basis for harmonic oscillator}
\label{krylovforho}

In this appendix, we construct the Krylov basis for 
the harmonic oscillator Hamiltonian in Eq.~(\ref{HO}).
The derivative of the Hamiltonian sets the zeroth-order basis
$|\theta_0\rangle$.
It is given by 
\be
 \dot{H} &=& -\dot{q}_0\sqrt{\frac{m\omega^3}{2}} (C^\dag+C)
 +\frac{\dot{\omega}}{2}\left(C^{\dag 2}+C^2\right)
 \no\\ &&
 +\frac{\dot{\omega}}{\omega}H.
\ee
Evaluating the commutator of the operators that appeared in $\dot{H}$,
we obtain 
\be
 && \mathcal{L}(C^\dag+C)=\omega(C^\dag-C), \\
 && \mathcal{L}\left(C^{\dag 2}+C^2\right)=2\omega\left(C^{\dag 2}-C^2\right).
\ee
These lead to new operators that can, however, be expressed in terms of the original operator set as 
\be
 && \mathcal{L}(C^\dag-C)=\omega(C^\dag+C), \\
 && \mathcal{L}\left(C^{\dag 2}-C^2\right)=2\omega\left(C^{\dag 2}+C^2\right).
\ee
Since the original Hamiltonian is real symmetric, the $L$ matrix
has the structure in Eq.~(\ref{LM}).
Using these results, we can set 
\be
 && X_1 = \frac{C^\dag C+\frac{1}{2}}
 {\sqrt{\langle\left(C^\dag C+\frac{1}{2}\right)^2\rangle}}, \\
 && X_2 = \frac{C^\dag +C}{\sqrt{\langle\left(C^\dag+C\right)^2\rangle}}, \\
 && X_3 = \frac{C^{\dag 2} +C^2}
 {\sqrt{\langle\left(C^{\dag 2}+C^2\right)^2\rangle}} 
\ee
and 
\be
 && Y_1 = i\frac{C^\dag -C}{\sqrt{-\langle\left(C^\dag-C\right)^2\rangle}}, \\
 && Y_2 = i\frac{C^{\dag 2} -C^2}
 {\sqrt{-\langle\left(C^{\dag 2}-C^2\right)^2\rangle}},
\ee
where $\langle\cdot\rangle$ denotes the average ${\rm Tr}[\rho(H)(\cdot)]$.
Since the dimension of the Hilbert space is infinite in the present system,
we need to choose the density operator $\rho(H)$ in the inner product
in a proper way.
For example, we can use the canonical Gibss--Boltzmann distribution
$\rho(H)= e^{-\beta H}/{\rm Tr}\,e^{-\beta H}$.

Now, we obtain 
\be
 M = i\omega\left(\begin{array}{cc}
 0 & 0 \\ 1 & 0 \\ 0 & 2 \end{array}\right).
\ee
The Krylov basis is constructed by choosing 
the initial normalized vector $|\theta_0\rangle=(x,y,z)^{\rm T}$.
We obtain 
\be
 && |\theta_1\rangle = \frac{-i}{\sqrt{y^2+4z^2}}\left(\begin{array}{c}
 y \\ 2z \end{array}\right), \\
 && |\theta_2\rangle = \frac{1}{\sqrt{y^2+16z^2-(y^2+4z^2)^2}}
 \left[\left(\begin{array}{c}
 0 \\ y \\ 4z \end{array}\right)\right. \no\\
 && \left.
 -(y^2+4z^2)\left(\begin{array}{c}
 x \\ y \\ z \end{array}\right)
 \right], \\
 && |\theta_3\rangle = -i\frac{yz}{|yz|}\frac{1}{\sqrt{y^2+4z^2}}
 \left(\begin{array}{c}
 -2z \\ y \end{array}\right), \\
 && |\theta_4\rangle = \frac{xyz}{|xyz|}
 \frac{1}{\sqrt{y^2+16z^2-(y^2+4z^2)^2}}
 \left(\begin{array}{c}
 3yz \\ -4zx \\ xy \end{array}\right). \no\\ 
\ee
The corresponding Lanczos coefficients are given by 
\be
 && b_1=\omega\sqrt{y^2+4z^2}, \\
 && b_2 = \omega\sqrt{\frac{y^2+16z^2-(y^2+4z^2)^2}{y^2+4z^2}}, \\
 && b_3 = \frac{6\omega|yz|}{\sqrt{(y^2+4z^2)[y^2+16z^2-(y^2+4z^2)^2]}}, \\
 && b_4 = 2\omega|x|\sqrt{\frac{y^2+4z^2}{y^2+16z^2-(y^2+4z^2)^2}}.
\ee
The expansion terminates at the fifth order, which means
$d=5$ and $d_A=2$ assuming $x$, $y$, and $z$ are nonzero.
As we discuss in the main body of the paper, 
the condition $\dot{q}_0=0$ gives $y=0$, and 
we obtain $d=3$ and $d_A=1$. For $\dot{\omega}=0$,
one finds $x=z=0$  and obtains that $d=2$ and $d_A=1$.
 
\section{Krylov basis for the one-dimensional transverse-field Ising model}
\label{krylovforqising}

For the Hamiltonian in Eq.~(\ref{Hqising}), we apply the Krylov algorithm to find 
\be
 && O_0 = -M, \\
 && b_0=\frac{\sqrt{n_s}}{2}v\dot{g}, \\
 && O_1 = -iW_1 , \\
 && b_1=\sqrt{2}v, \\
 && b_2O_2=\sqrt{2}v\left[-V_2^X+g(V_1^X-V_1^Y)\right], \\
 && b_2=\sqrt{2}v\sqrt{1+2g^2}, \\
 && O_3=iW_2, \\
 && b_2b_3=4v^2g.
\ee
This result implies that $O_{2k-1}=(-1)^k iW_k$ at odd order.
Assuming that the relation holds for $O_{2k-1}$, 
we calculate $\mathcal{L}^2O_{2k-1}$ to find 
\be
 && b_{2k}b_{2k+1}O_{2k+1} = 4v^2g(-1)^{k+1}iW_{k+1} \no\\
 && +(-1)^ki\left[4v^2(1+g^2)-(b_{2k-1}^2+b_{2k}^2)\right]W_k.
\ee
Since this operator is orthogonal to $W_k$, 
$b_{2k-1}^2+b_{2k}^2=4v^2(1+g^2)$ must be satisfied.
Then, from the normalization condition, we obtain 
$O_{2k+1}=(-1)^{k+1} iW_{k+1}$ and $b_{2k}b_{2k+1}=4v^2g$.

For a given set of Lanczos coefficients, $\alpha_k$ is obtained by 
solving Eq.~(\ref{alphaoddd}).
Here, we denote the matrix in the equation by $K$ and 
represent its spectral decomposition as 
$K=\sum_{k=1}^{d_A} \lambda_k |\phi_k\rangle\langle \phi_k|$.
The eigenvalues and eigenstates are explicitly obtained as 
\be
 \lambda_k&=&4v^2\left(1+g^2-2g\cos\frac{\pi k}{d_A+1}\right), \\
 |\phi_k\rangle &=& \sqrt{\frac{2}{d_A+1}}\left(\begin{array}{cc}
 \sin\frac{\pi k}{d_A+1} \\ -\sin\frac{2\pi k}{d_A+1} \\ \vdots \\
 (-1)^{d_A-1}\sin\frac{d_A\pi k}{d_A+1} \end{array}\right).
\ee
Then, we can write 
\be
 \alpha_k = \sum_{l=1}^{d_A}\frac{-b_1}{\lambda_l}
 \langle k|\phi_l\rangle\langle\phi_l|1\rangle
\ee
and find Eq.~(\ref{aqising}).

\bibliography{STAKrylov_lib}

\end{document}